\documentclass{aa}
\usepackage{graphicx}
\usepackage{txfonts}
\usepackage{longtable,lscape}
\usepackage{natbib}
\bibpunct{(}{)}{;}{a}{}{,} 
\begin{document}
   \title{Combined Chandra, XMM-Newton and Hubble Space Telescope observations of the Galactic globular cluster \mbox{\object{NGC 2808}}}


   \author{M. Servillat\inst{1}
          \and
          A. Dieball\inst{2}
          \and
          N. A. Webb\inst{1}
          \and
          C. Knigge\inst{2}
          \and
          R. Cornelisse\inst{3}
          \and \\
          D. Barret\inst{1}
          \and
          K. S. Long\inst{4}
          \and
          M. M. Shara\inst{5}
          \and
          D. R. Zurek\inst{5}
          }

   \institute{CESR, Observatoire Midi-Pyr\'en\'ees, Universit\'e Paul Sabatier, CNRS, 9 avenue du
              Colonel Roche, BP 44346, 31028 Toulouse Cedex 4, France --
              \email{mathieu.servillat@cesr.fr}
              \and
              Department of Physics and Astronomy, University of
              Southampton, SO17 1BJ, UK
              \and
              Instituto de Astrofisica de Canarias, Via Lactea, La Laguna E-38200, Santa Cruz de Tenerife, Spain
              \and
              Space Telescope Science Institute, Baltimore, MD 21218, USA
              \and
              Department of Astrophysics, American Museum of Natural
              History, New York, NY 10024, USA
             }

   \date{Received -; accepted -}


  \abstract
  {}
   {Using new Chandra X-ray observations and existing XMM-Newton X-ray and Hubble far ultraviolet observations, we aim to detect and identify the faint X-ray sources belonging to the Galactic globular cluster \mbox{\object{NGC 2808}} in order to understand their role in the evolution of globular clusters.}
   {We present a Chandra X-ray observation of the Galactic globular cluster \mbox{\object{NGC 2808}}. We classify the X-ray sources associated with the cluster by analysing their colours and variability. Previous observations with XMM-Newton and far ultraviolet observations with the Hubble Space Telescope are re-investigated to help identify the Chandra sources associated with the cluster. We compare our results to population synthesis models and observations of other Galactic globular clusters.}
   {We detect 113 sources, of which 16 fall inside the half-mass radius of \mbox{\object{NGC 2808}} and are concentrated towards the cluster core. From statistical analysis, these 16 sources are very likely to be linked to the cluster. We detect short-term (1~day) variability in X-rays for 7 sources, of which 2 fall inside the half-mass radius, and long-term (28~months) variability for 10 further sources, of which 2 fall inside the half-mass radius.
   Ultraviolet counterparts are found for 8 Chandra sources in the core, of which 2 have good matching probabilities and have ultraviolet properties expected for cataclysmic variables.
   We find one likely neutron star-quiescent low-mass X-ray binary and 7 cataclysmic variable candidates in the core of \mbox{\object{NGC 2808}}. The other 8 sources are cataclysmic variable candidates, but some could possibly be active binaries or millisecond pulsars.
   We find a possible deficit of X-ray sources compared to \mbox{\object{47 Tuc}} which could be related to the metallicity content and the complexity of the evolution of \mbox{\object{NGC 2808}}.}
  {}

   \keywords{Galaxy: globular clusters: individual: \mbox{\object{NGC 2808}} --
             X-rays: general --
             Stars: binaries: close --
             Cataclysmic variables
            }

   \titlerunning{\textit{Chandra}, \textit{XMM-Newton} and \textit{HST} observations of \mbox{\object{NGC 2808}}}
   \authorrunning{Servillat et al.}
   \maketitle
%

\section{Introduction}

Globular clusters (GCs) are old, gravitationally bound stellar
systems which can have extremely high stellar densities, especially in
their core regions. In such an environment, dynamical
interactions between the cluster members are inevitable, leading to
a variety of close binary (CB) systems and other exotic stellar
objects. The observed overabundance of neutron star (NS) low-mass X-ray
binaries (LMXBs) in GCs relative to the Galactic field was explained
by the dynamical processes
occurring in the dense cores of GCs \citep{fabian}.
In contrast, evolution of a primordial binary into an
LMXB in a GC is considered to be much less likely
\citep{VH87}.
Observations also support the fact that quiescent LMXBs
(qLMXBs) in GCs scale with the cluster encounter rate
\citep{GBW03,Pooley+03}, implying that qLMXBs are formed through
dynamical processes in the dense cores.
As white dwarfs (WDs) are far
more common than NSs, we would then also expect many more close
binaries containing an accreting WD primary,
i.e. cataclysmic variables (CVs).

The dynamically-formed CBs are expected to be found in the cores of GCs,
where the stellar densities are at a maximum. The
less dense regions outside the cores might be populated by CBs that
evolved from primordial binaries \citep[e.g.][]{Davies97} which are unlikely
to survive in the dense core region. \citet{HAS07} found that the
combined effects of new binary creation and mass segregation exceed
the destruction of primordial binaries in the central region of GCs,
leading to a marked increase of the binary fraction in the central
regions. Thus, we expect the majority of CBs,
which are more massive than the mean stellar mass, to be located
inside the half-mass radius. Outside the half-mass radius, the
primordial binary fraction is well preserved \citep{HAS07}.

CBs are important for our understanding of GC evolution, since the
binding energy of a few, very close binaries can rival that of a
modest-sized GC \citep[e.g.][and references
  therein]{EHI87,Hut+92,Hut+03}. In the core,
binaries are subject to encounters and hard binaries become harder
while transferring their energy to passing stars. Thus, CBs can
significantly affect the dynamical evolution of the cluster. If there are only a few CBs,
thermal processes dominate the cluster evolution, leading to core
collapse followed by GC disruption on a timescale shorter than the
mean age of GCs, estimated to be $12.9\pm2.9$~Gyr \citep{CGCFP00}. In
contrast, the presence of many CBs leads to violent interactions,
which heat the cluster, delay the core collapse, and promote its
expansion. This depends
critically on the number of CBs, which is still poorly known.

Finding and studying these systems has
proven to be extremely difficult, since the spatial resolution and
detection limits of most available telescopes are insufficient for their
detection. Only with the improved sensitivity and imaging quality of
XMM-Newton and Chandra in the X-ray \citep[e.g.][]{Webb+04,WWB06,Heinke+03,Heinke+06} and HST in
the ultraviolet (UV) to infrared (IR) wavebands \citep[ and references therein]{GHEM01,Albrow+01,EGHG03,EGHG03b,Knigge+02,Knigge+03} has it become
possible to finally detect significant numbers of CB
systems in GCs.

The 13 bright X-ray
sources found in the $\sim$150 known Galactic GCs are LMXBs showing type I X-ray bursts
\mbox{\citep[e.g.][]{1983adsx.conf...41L}}, whereas the faint sources
belonging to the clusters are qLMXBs, CVs,
active binaries (ABs, generally RS~CVn systems), or millisecond pulsars (MSPs).
Multiwavelength studies can be used to identify the faint X-ray sources.
For example, qLMXBs are usually identified by their
soft blackbody-like or hydrogen atmosphere X-ray spectra \mbox{\citep[e.g.][]{GBW03,GBW03b}},
CVs can be confirmed by their blue, variable optical counterpart with hydrogen emission lines in their spectra
\mbox{\citep[e.g.][]{Webb+04}}, ABs by their main-sequence, variable
optical counterparts \mbox{\citep[e.g.][]{EGHG03}}, and MSPs by their
radio counterpart \mbox{\citep[e.g.][]{GHEM01}}.

Here, we present an X-ray study of the massive (\mbox{$\sim10^{6}
M_{\sun}$})
GC \mbox{\object{NGC 2808}} (\mbox{$\alpha = 09^{h}
12^{m} 02^{s}$, $\delta = $-$64^{\circ} 51{\arcmin} 47{\arcsec}$}). This
intermediate metallicity GC \citep[\mbox{$\rm{[Fe/H]} = -1.36$},
][]{Walker99} lies at a distance of 9.6~kpc and is reddened by
$E_{B-V}=0.22\pm0.01$ \citep{Harris96}. An absorption column of N${_H
  = 1.2\times10^{21}\mathrm{~cm^{-2}}}$ is derived from the reddening
with the relation computed by \citet{PS95}. The cluster has a very
dense and compact core ($0.26\arcmin$), a half-mass radius of
$0.76\arcmin$, a tidal radius of $15.55\arcmin$, and a half-mass
relaxation time of $1.35\times10^{9}$~yrs \citep{Harris96}.

\mbox{\object{NGC 2808}} has received considerable attention in the
literature and has been observed in the optical in detail as this GC
is one of the most extreme examples with an unusual horizontal branch
(HB) morphology, as first noted by \citet{Harris74}.
It shows a bimodal HB and one
of the longest blue HB tails, the so-called extreme HB (EHB),
with prominent gaps between the red HB (RHB), blue HB (BHB) and EHB
\citep[see also ][]{Bedin+00,Carretta+06}.
Recently, \citet{Piotto+07} found that \mbox{\object{NGC 2808}}'s main sequence (MS) is separated
into three branches, which might be associated with the complex HB
morphology and abundance distribution, and might be due to successive
rounds of star formation with different helium abundances.
\mbox{\object{NGC 2808}} is proposed as a good candidate to harbour an intermediate mass black hole (IMBH) in its core, due to its optical luminosity profile and EHB morphology \citep{Miocchi07}.

The core of \mbox{\object{NGC 2808}} has been imaged with the Space
Telescope Imaging Spectrograph (STIS) on board the Hubble Space
Telescope (HST) in the far-UV (FUV) and the near-UV (NUV).
\citet{Dieball+05} used the data set with an emphasis on the
dynamically-formed stellar populations like CVs and blue stragglers
(BSs) and young WDs. They found $\sim$40~WD, $\sim$60~BS
and $\sim$60~CV candidates in the field of view that covers the core
of the cluster.
Two of the CV candidates are variable (FUV sources~222 and 397).

\mbox{\object{NGC 2808}} has also been observed with XMM-Newton in Feb.
2005. \citet{Servillat+08-a} found 96~sources in the field of view (equivalent to the tidal radius), of
which five fall inside the half-mass radius and are likely to be linked to the cluster. One qLMXB candidate
and four CV candidates were discovered in the core of
\mbox{\object{NGC 2808}}. However, several sources remained unresolved.

In Sect.~\ref{obs}, we present the new Chandra X-ray data, and then compare them to
the XMM-Newton observations (Sect.~\ref{xmm}).
We present HST and XMM-Newton Optical Monitor UV counterparts in Sect.~\ref{uv}.
We finally discuss our results in
Sect.~\ref{disc}.


\section{X-ray observations}
\label{obs}

\mbox{\object{NGC 2808}} was observed with the Chandra X-ray
Observatory and the Chandra Advanced CCD Imaging Spectrometer-Imager
(\mbox{ACIS-I}) at its focus on 2007 June 19--21 (28 months after the XMM-Newton observation) for two distinct
exposures of 46 and 11~kilo seconds~(ks). The four front-illuminated \mbox{ACIS-I}
chips were used as well as the front-illuminated \mbox{ACIS-S2} chip and
the back-illuminated \mbox{ACIS-S3} chip at the edge of the field of
view. The data were taken in faint, timed-exposure mode.
The core of \mbox{\object{NGC 2808}} 
falls inside the \mbox{ACIS-I3} chip, where the highest resolution is
achieved.

\subsection{Data reduction}

Data reduction was performed using the CIAO~v3.4
software\footnote{http://cxc.harvard.edu/ciao} \citep{Fruscione+06} and the CALDB
v3.4.0 set of calibration files (gain maps, quantum efficiency, quantum
efficiency uniformity, effective area).
We reprocessed the level 1 event files of both observations without
including the pixel randomization that is added during standard
processing. This method slightly improves the point-spread function (PSF).
We removed cosmic-ray events which could be detected as spurious faint
sources using the tool \textit{acis\_detect\_afterglow}, and
identified bad pixels with the tool \textit{acis\_run\_hotpix}. The
event lists were then filtered for grades, status, and good time
intervals (GTIs), as it is done in the standard processing. We selected
events within the energy range \mbox{0.3--10.0~keV}, where good sensitivity is achieved.

The two observations are successive, the fields of view and the roll
angles are similar, and the PSFs from one
observation to another are also similar. Therefore the two epochs can be
processed as a single observation. We thus reprojected the event list
of the second observation to match the first observation, and merged
the two event lists using the thread \textit{merge\_all}. We generated
an exposure map of the field of view using the same thread.
This led to 56.9~ks of clean observation (sum of GTIs).

\subsection{Source detection}
\label{sec:sourcedetect}

\begin{figure*}
\centering
\includegraphics[width=\textwidth]{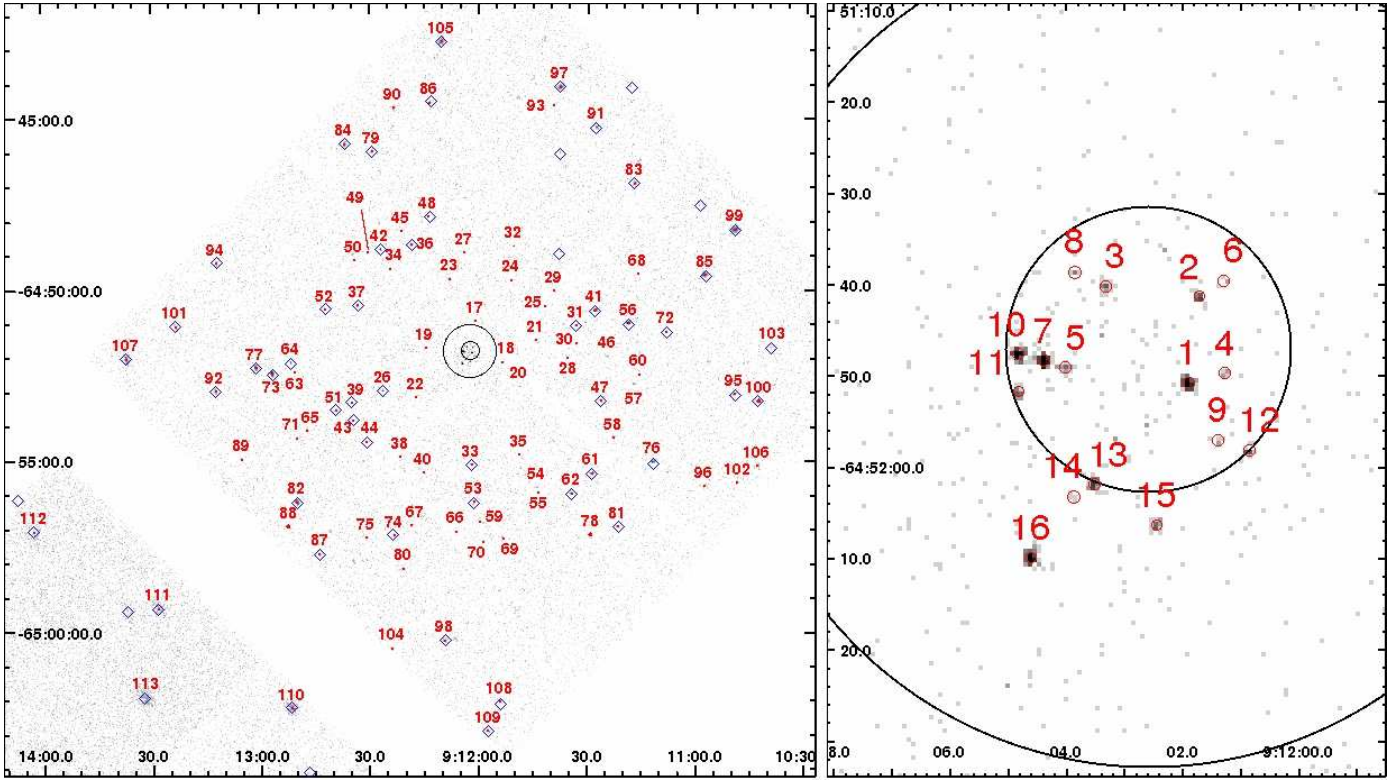}
\caption{Chandra source~map in the direction of \mbox{\object{NGC 2808}}. Events are
  selected in the energy band 0.5--6.0~keV where the signal to noise
  ratio is the highest. Core and half-mass radii are
  shown. Chandra $1\sigma$ positional error circles are represented. \textit{Left}: Smoothed Chandra image of the field, XMM-Newton sources are
  shown as diamonds. \textit{Right}: Zoom of the Chandra core region of
  \mbox{\object{NGC 2808}}, XMM-Newton sources are not shown here for clarity (see Fig. \ref{fig:n2808_core} for a comparison of XMM-Newton and Chandra sources in the core).
	}
	\label{fig:chandra_map}
\end{figure*}

In order to obtain a list of source candidates, we employed the CIAO wavelet-based \textit{wavdetect} tool for source detection in the field of view covered by the four \mbox{ACIS-I} chips.
Two energy bands were used, the 0.3--10.0~keV band with all events which allows the detection of the faintest sources, and 0.5--6.0~keV with a higher signal to noise ratio which gives secure detections.
We selected scales of 1.0, 1.4, 2.0, 2.8, 4.0, and 5.6 pixels. The scales were chosen to look for narrow PSF sources on-axis and ensure optimal separation, and larger PSF sources at the edge of the detectors, where the PSF is degraded.
We selected a threshold probability of $10^{-6}$, designed to give one false source~per $10^{6}$ pixels.
This led to the detection of 88 source~candidates of which $\sim6$ may be false. Of these sources, eleven fall inside the half-mass radius of \mbox{\object{NGC 2808}}.

To this list, we added other source~candidates in order to
test their significance in further processing. We added 14 X-ray sources previously detected with XMM-Newton
\citep{Servillat+08-a}.
Seven of them fall on the two \mbox{ACIS-S} chips
where we did not use \textit{wavdetect} as the PSF is too large, and seven others were not detected by
\textit{wavdetect} on the \mbox{ACIS-I} chips and have possibly varied in flux.
Several other faint sources can be
picked out by eye, but are not identified by \textit{wavdetect}, possibly because of crowding on-axis and lower signal to noise at the edge of the detector due to the vignetting.
We thus added 22 source~candidates, of which 10 are inside the half-mass
radius of \mbox{\object{NGC 2808}}. In total we listed 124 source candidates.

We then used
ACIS~Extract\footnote{http://www.astro.psu.edu/xray/docs/TARA/ae\_users\_guide.html}
\citep{acisextract} to refine the positions of the sources and to estimate the significance of each candiate source.
We excluded sources with a probability of
being a source~lower than 99.99\%. This probability is estimated from
the number of source candidate counts compared to the
surrounding background counts, taking into account Poisson statistics
\citep[][ \S5.9.3]{acisextract}.
Our final list has 113 sources, of which 16 are
located inside the half-mass radius of \mbox{\object{NGC 2808}}. All of these sources have more
than four counts. They are
numbered according to their offset from the center of NGC 2808. Their
properties are listed in Table~\ref{tab:chandra_sources}, and a source
map is presented in Fig.~\ref{fig:chandra_map}. The core image is compared to the
XMM-Newton image of \citet{Servillat+08-a} in Sect.~\ref{sec:Chandra_vs_XMM}.
From the faintest sources detected, the limiting flux of the observation is ${L_{[0.5-8keV]}\sim0.9\times10^{-15}\mathrm{~erg~cm^{-2}~s^{-1}}}$, corresponding to a limiting luminosity of ${L_{[0.5-8keV]}\sim1.0\times10^{31}\mathrm{~erg~s^{-1}}}$ (at the distance of the cluster).

\addtocounter{table}{1}

The $1\sigma$ position error reported in Table~\ref{tab:chandra_sources} is used in this paper to look for matching XMM-Newton and HST FUV sources.
One may estimate the absolute position error by adding in quadrature the pointing accuracy of Chandra ($0.4\arcsec$, $1\sigma$ error) and the spatial distortion error over the detectors ($0.1\arcsec$, $1\sigma$ error).

\subsection{Members of \mbox{\object{NGC 2808}}}

\begin{table}[b]
\caption{Expected background sources and detected X-ray sources in \mbox{\object{NGC 2808}}
  field of view. Results are presented assuming a detection limit and a completeness limit (see text).}             
\label{tab:lnls}      
\centering                          
\begin{tabular}{r@{ -- }l|cc|cc}        
\hline\hline                 
\multicolumn{2}{c|}{~}   & \multicolumn{2}{c|}{Detection limit} & \multicolumn{2}{c}{Completeness limit} \\    
\multicolumn{2}{c|}{Annulus}   & Expected & Detected & Expected & Detected \\    
\hline                        
   0\arcmin   & 1.8\arcmin  & $ 4.0\pm0.8$ & 20 & $ 3.4\pm0.8$ & 12  \\
   1.8\arcmin & 3.8\arcmin  & $12.7\pm2.0$ & 20 & $ 9.8\pm2.0$ & 11  \\
   3.8\arcmin & 7.0\arcmin  & $36.2\pm6.0$ & 43 & $27.2\pm6.0$ & 30  \\
\hline                        
   0\arcmin   & 0.76\arcmin & $ 0.8\pm0.8$ & 16 & $ 0.6\pm0.8$ & 11  \\
\hline                        
\end{tabular}
\end{table}

We estimated the number of background X-ray sources we expect to
detect in our observation in order to deduce the number of sources
likely to be linked to the cluster.

We divided the field of view into three annuli to account for
vignetting, and to include in each region at least 20 detected
sources.
We used the $log(N)$--$log(S)$ relation calculated by \citet{HMS05} \citep[see also][]{Giacconi+01,Hasinger+01} to
convert in each annulus our minimum detectable fluxes ($S$) into the number of background
sources expected ($N$). This relation was derived from a
survey of soft X-ray active galactic nuclei (AGN) in the energy range
0.5--2.0~keV performed with both XMM-Newton and Chandra.
We took into account two errors on the value read
from the $log(N)$--$log(S)$ diagram which were added in quadrature: the error on the Chandra flux estimate (see Table~\ref{tab:chandra_sources}) converted to an error in $N$ while reading the $log(N)$--$log(S)$ diagram, and the precision of the relation which includes Poisson noise \citep[see error bars in][their Fig.~3]{Giacconi+01}.
For each annulus, we estimated the minimum detectable
unabsorbed flux of a point source in the energy range 0.5--2.0~keV
using WebPIMMS\footnote{http://heasarc.gsfc.nasa.gov/Tools/w3pimms.html}~v3.9b \citep{Mukai93}.
We assumed for the source~a power law model of photon index 1.5 (mean
of the detected sources) and the absorption of the cluster.
We assumed a minimum detectable count rate corresponding to two cases: the detection limit (the count rate of the faintest source in each annulus) and the completeness limit (twice the count rate of the faintest source). These values were corrected for the vignetting in each annulus using the exposure map.
The estimates might be slightly overestimated for the detection limit as we have an incomplete sample of sources.
The results are shown in Table~\ref{tab:lnls}.

\begin{figure}
\centering
\includegraphics[width=\columnwidth]{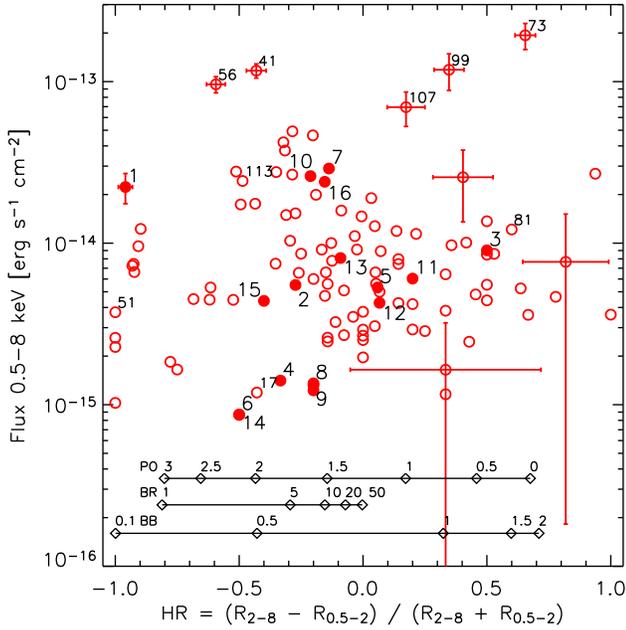}
\caption{Colour-flux diagram of Chandra sources in the direction of
  \mbox{\object{NGC 2808}}. For clarity, we labelled only the core sources (filled
  circles), the brightest sources, and variable sources (see Fig.~\ref{fig:chandra_lcvar}). The values for all sources can be found in Table~\ref{tab:chandra_sources}. Some error bars are
  shown which are representative of the error bars at the same flux. Black lines with diamonds show the colours (with an arbitrary flux) of different models with an absorption of $1.2\times10^{21}\mathrm{~cm^{-2}}$: \textbf{PO}:~power law with photon indices 3, 2.5, 2, 1.5, 1,
0.5, 0. \textbf{BR}:~thermal bremsstrahlung with temperatures
1, 5, 10, 15, 20, 50~keV. \textbf{BB}:~blackbody spectrum
with temperatures 0.1, 0.5, 1, 1.5, 2~keV.
	}
	\label{fig:chandra_cmd}
\end{figure}

\begin{figure}
\centering
\includegraphics[width=\columnwidth]{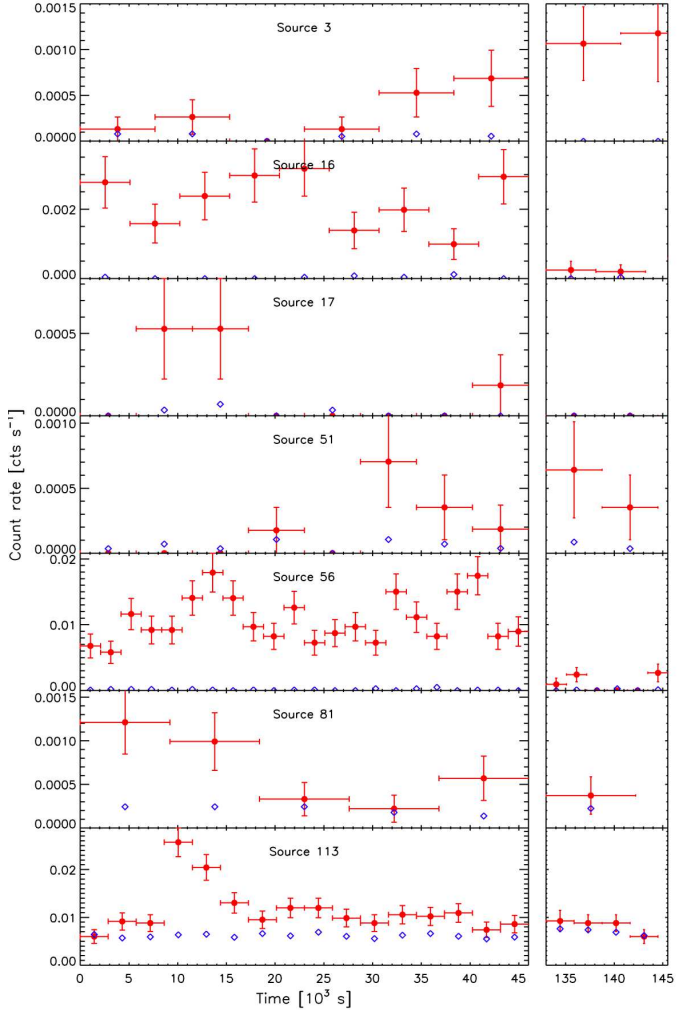}
\caption{Lightcurves of variable Chandra sources. The data is shown with filled circles and error bars, and the corresponding background extracted in an annulus around the source is shown with diamonds. The first (left) and second (right) observations are separated by 86~ks. The zero time is 2007 June 19 at 15\,h\,55\,min\,53.5\,s.
	}
	\label{fig:chandra_lcvar}
\end{figure}

An excess is clearly seen in the center. In the annulus 1.8--3.8\arcmin, an excess is seen if we assume the detection limit (Table~\ref{tab:lnls}), however, with the completeness limit this excess is not confirmed.
In the last annulus, we detect the expected number of background sources within the errors.
We performed the same estimation inside the half-mass radius of \mbox{\object{NGC 2808}} and $0.8\pm0.8$ sources are expected.
Therefore,
the 16 sources inside the half-mass radius are very likely to be cluster sources, with perhaps one background source aligned fortuitously.
As \mbox{\object{NGC 2808}} has a low Galactic latitude \citep[$b=-11.3\degr$,][]{Harris96},
we may also expect foreground sources such as active stars or field CVs.

\subsection{Spectral and variability analysis}

We used the ACIS~Extract procedure \textit{ae\_standard\_extraction} to
extract spectra and lightcurves for each source. We estimated for
each source an optimal extraction region to enclose 90\% of the PSF,
and estimated the background by selecting 100~surrounding counts
outside source regions.
None of the extraction regions overlap.
The fluxes are then estimated from the count
rates in several bands, and the \mbox{0.5--8~keV} flux is
derived. Hardness ratios (colours) are estimated from two energy bands, 0.5--2
and 2--8~keV. The response files of the detector were computed with ACIS~Extract for
each source using the CIAO tasks \textit{mkarf} and \textit{mkacisrmf} with the associated gain files.
We performed a generic spectral fitting of the sources with a power law
model and the absorption of the cluster using Xspec~v12.2
\citep{Arnaud96} through the procedure \textit{acis\_extract}
(FIT\_SPECTRA stage). The unabsorbed flux is estimated from the best
fit of this model. All
these results are reported in Table~\ref{tab:chandra_sources}, and we
show a colour-flux diagram of all the sources in
Fig.~\ref{fig:chandra_cmd}.

\begin{figure*}[t]
\centering
\includegraphics[width=\textwidth]{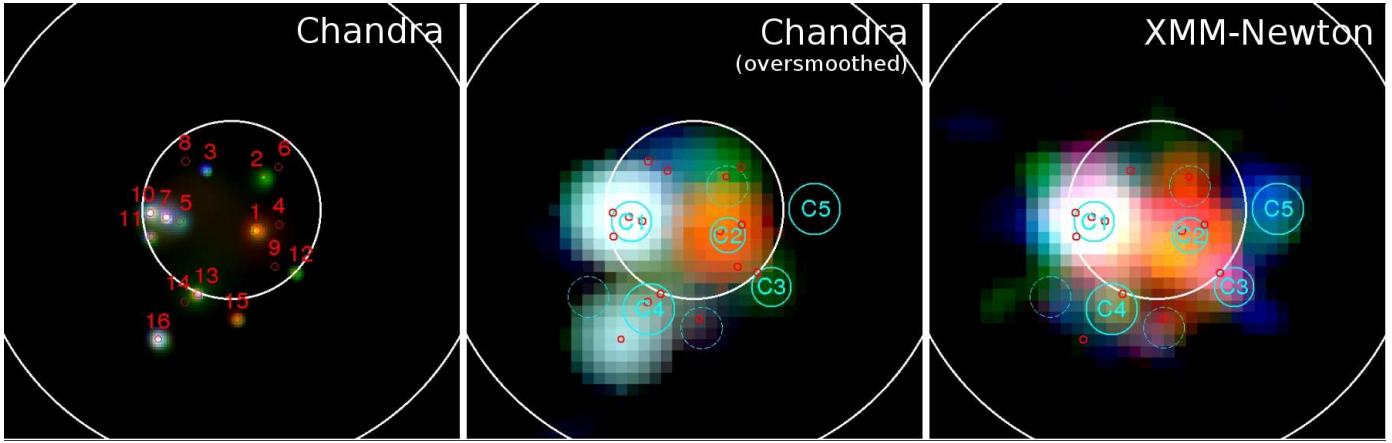}
\caption{Images of the core of \mbox{\object{NGC 2808}}. Colours correspond to
  different energy bands, red:~0.5--1.5~keV, green:~1.5--3~keV,
  blue:~3--8~keV. The absolute $1\sigma$ positional error for each source is represented as a circle, red (small) for Chandra and blue (large) for XMM-Newton. Core and half-mass radii are shown.
  \textit{Left}:~Chandra image, smoothed using the adaptative smooth tool \textit{csmooth}.
  \textit{Center}:~Chandra image, over smoothed with a Gaussian filter to be compared with XMM-Newton image.  \textit{Right}:~XMM-Newton combined image (PN, MOS1 and MOS2), smoothed with a Gaussian filter. Only Chandra detected sources which could have been detected by XMM-Newton are represented (small red circles).
	}
	\label{fig:n2808_core}
\end{figure*}

A Kolmogorov-Smirnov (KS) test was performed on
the extracted and unbinned lightcurve in order to detect variable sources.
Seven sources are found to have a KS probability of constancy lower than $10^{-2}$, of which two are
located in the core of \mbox{\object{NGC 2808}} (sources 3 and 16), and we confirmed the variability of these sources using Poisson statistics.
We extracted their lightcurves and the background lightcurves in an annulus around the source with CIAO \textit{dmextract} task. The binned lightcurves are shown in Fig.~\ref{fig:chandra_lcvar}.

\section{Comparison with XMM-Newton observations}
\label{xmm}

The XMM-Newton X-ray observation of \mbox{\object{NGC 2808}}, performed on 2005
February $1^{st}$ \citep{Servillat+08-a}, was reprocessed in order to
match our Chandra energy bands (0.5--2 and \mbox{2--8}~keV). We used the
XMM-Newton Science Analysis System\footnote{http://xmm.vilspa.esa.es/sas} (SAS v7.1) and the most
recent calibration data files. The data reduction is detailed in
\citet{Servillat+08-a}. The 96 detected sources in the XMM-Newton
field of view were reprocessed with \textit{emldetect}, without
refining the position, in order to extract fluxes and hardness
ratios. These values are reported in Table~\ref{tab:chandra_sources}
for the sources that are inside the Chandra field of view.
We took into account the different sensitivities of the
instruments in the energy bands used.
Using a power law model of photon indices 0, 1, 2 or 3 and the absorption of the cluster, we converted with WebPIMMS a given flux into Chandra ACIS-I and XMM-Newton pn count rates, and compared the hardness ratios.
The following conversion factors from XMM-Newton pn to Chandra ACIS-I count rates were deduced: C$_{0.5-2}=0.75$ and C$_{2-8}=1.00$. The corrected hardness ratios for the different models used are found to match with a maximum error of $0.07$, comparable to the $1\sigma$ errors on hardness ratios (see Table~\ref{tab:chandra_sources}).
Hereafter we refer to the corrected hardness ratios for XMM-Newton sources.

\subsection{Inside the half-mass radius}
\label{sec:Chandra_vs_XMM}

\begin{figure}
\centering
\includegraphics[width=\columnwidth]{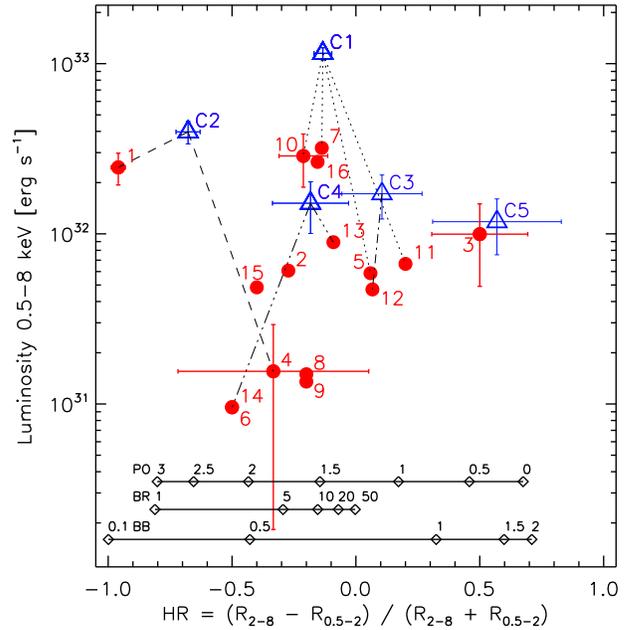}
\caption{Colour-luminosity diagram of Chandra and XMM-Newton core
  sources in \mbox{\object{NGC 2808}}. Filled red circles represent Chandra sources,
  and open blue triangles XMM-Newton sources. XMM-Newton sources are resolved
  into several Chandra sources which are linked with a line. For
  clarity, only a few error bars are shown. Black lines correspond to the models described in Fig.~\ref{fig:chandra_cmd}.
	}
	\label{fig:chandra_xmm_cmd_core}
\end{figure}

The better resolution of Chandra allowed us to resolve the core
sources previously detected with XMM-Newton. At first, in order to
compare these different observations, we degraded the Chandra image
with a Gaussian filter to enlarge the PSF to the XMM-Newton PSF
size. The result is presented in Fig.~\ref{fig:n2808_core}. We note
that the images are in general very similar. XMM-Newton source~C5 is
clearly missing in the Chandra observation, and Chandra source~16 was
not detected with XMM-Newton. We extracted a spectrum and generated response files at the position
of Chandra source~16 in the XMM-Newton observation to estimate a flux
detection limit for this source.
We found
that Chandra source~16 has varied by a factor of at least~$\sim5$.
XMM-Newton source~C5 has also varied by a factor of at least~$\sim5$.

We show a colour-luminosity diagram for Chandra core sources with corresponding XMM-Newton core
sources in Fig.~\ref{fig:chandra_xmm_cmd_core}. We note that the observations are
well correlated, taking into account the different resolutions.
XMM-Newton source~C1 is resolved into four Chandra sources (5, 7, 10 and 11) with
consistent colours.
XMM-Newton source~C2 matches with Chandra source~1. This source appeared slightly harder than the Chandra source, probably because it is overlapped by XMM-Newton source~C1 which is hard, as shown in
\citet{Servillat+08-a}.
XMM-Newton source~C3 is matching with Chandra source~12, and XMM-Newton source~C4 has consistent parameters with Chandra source~13.
We also note that Chandra sources 2 and 15 are related to $2\sigma$ detections in the XMM-Newton observation, and their fluxes are consistent with the detection limit of the latter observation \citep{Servillat+08-a}.

We extracted spectra and response files for XMM-Newton core sources C1, C3 and C4, and fitted with Xspec a power law with the absorption of the cluster, and a small contribution  from close sources as the PSFs overlap. We also extracted a combined spectrum for Chandra sources 5, 7, 10 and 11, and spectra for Chandra sources 12 and 13, and performed a similar fitting. The results are presented in Table~\ref{tab:xmmch_core}. XMM-Newton source~C3 is more luminous than its matching Chandra source (12, $2.3\sigma$ variation).

\begin{table}
\caption{Fit results for XMM-Newton core sources C1, C3 and C4, and corresponding Chandra sources or association of sources. The fit was performed with a power law model with the absorption in the direction of the cluster. The photon index ($\Gamma$) and the unabsorbed flux in the 0.5--8~keV energy band in [${\times10^{-14}\mathrm{erg~cm^{-2}~s^{-1}}}$] are given.}     
\label{tab:xmmch_core}      
\centering                          
\begin{tabular}{ccc}        
\hline\hline                 
XMM ID & $\Gamma$ & Flux \\
\hline                        
C1 & $1.41\pm0.12$ & $7.3\pm0.7$ \\
C3 & $1.36\pm0.20$ & $2.1\pm0.5$ \\
C4 & $1.73\pm0.30$ & $1.1\pm0.5$ \\
\hline\hline                        
Chandra ID & $\Gamma$ & Flux \\    
\hline                        
5, 7, 10, 11 & $1.39\pm0.15$ & $5.7\pm1.2$ \\
   12        & $1.33\pm0.85$ & $0.4\pm0.2$ \\
   13        & $1.48\pm0.50$ & $0.8\pm0.3$ \\
\hline                        
\end{tabular}
\end{table}

\subsection{Other sources}

Overall, the sources are found to have consistent fluxes and colours in both observations (see Table~\ref{tab:chandra_sources}).
Chandra source~99 has faded by a factor 5 and has become
harder. Chandra sources 31, 73 and 86 appear softer, and 109 and
111 harder than in the XMM-Newton observation ($3\sigma$ variations).

\label{sec:xmmundet}
For the seven XMM-Newton sources outside the half-mass radius and XMM-Newton source C5, that were not detected with Chandra, we estimated the count rate expected with Chandra.
Then, we converted these count rates into counts according to the exposure map of our Chandra observation, i.e. correcting for the vignetting.
We estimated a detection threshold by looking at the number of counts of the faintest source detected with the same vignetting.
The results are given in Table~\ref{tab:xmmonly}.
We conclude that XMM-Newton sources~34, 66 and C5 should have been detected, unless they have varied between XMM-Newton and Chandra observations.
In particular, C5 must have varied by a factor of at least 5 in flux.

\begin{table}
\caption{XMM-Newton sources undetected with Chandra. The XMM-Newton ID, the flux in ${\times10^{-14}\mathrm{erg~cm^{-2}~s^{-1}}}$ and the hardness ratio are shown for each source. We give the Chandra chip type where the source was expected to be detected, the expected number of counts, and the detection threshold in counts. The expected counts that are significantly higher than the threshold are in bold face.}     
\label{tab:xmmonly}      
\centering                          
\begin{tabular}{@{~}c@{~~}c@{~~~}cc@{~~}c@{~~}c@{~}}        
\hline\hline                 
XMM ID & XMM flux & XMM HR & Chip & Exp. & Thresh. \\    
\hline                        
   34 &  1.31 $\pm$ 0.44 &     $-$0.39 $\pm$ 0.13 & ACIS-I &  \textbf{70 $\pm$  8} &  20 $\pm$  4  \\
   48 &  2.01 $\pm$ 0.69 &     $-$0.06 $\pm$ 0.15 & ACIS-S & 115 $\pm$ 10 & 150 $\pm$ 12  \\
   65 &  1.04 $\pm$ 0.47 &     $-$0.93 $\pm$ 0.14 & ACIS-S &  51 $\pm$  7 & 150 $\pm$ 12  \\
   66 &  0.36 $\pm$ 0.22 &     $-$0.67 $\pm$ 0.20 & ACIS-I &  \textbf{21 $\pm$  4} &   6 $\pm$  2  \\
   69 &  0.66 $\pm$ 0.35 &     $-$0.30 $\pm$ 0.25 & ACIS-I &  29 $\pm$  5 &  25 $\pm$  5  \\
   72 &  0.78 $\pm$ 0.46 &     $-$0.79 $\pm$ 0.21 & ACIS-S &  74 $\pm$  8 & 150 $\pm$ 12  \\
   84 &  0.27 $\pm$ 0.28 &     $-$0.38 $\pm$ 1.45 & ACIS-I &  10 $\pm$  3 &  30 $\pm$  5  \\
   C5 &  1.07 $\pm$ 0.38 &      0.57 $\pm$ 0.26 & ACIS-I &  \textbf{28 $\pm$  5} &   4 $\pm$  2  \\
\hline                        
\end{tabular}
\end{table}


\section{Counterparts to the X-ray sources}
\label{uv}

\subsection{Ultraviolet counterparts in the core}
\label{sec:uv_core}

The core of \mbox{\object{NGC 2808}} has been observed with the Space Telescope Imaging
Spectrograph (STIS) on board the {\it HST} in January/February 2000
using the F25QTZ filter, centered at 159~nm in the FUV band, and
the F25CN270 filter, centered at 270~nm in the NUV band.
The mosaic of FUV images has a radius of $\sim50\arcsec$\ and
only covers the core region of \mbox{\object{NGC 2808}}.

We searched for FUV counterparts to our Chandra sources, using the FUV
catalogue provided by \citet{Dieball+05}. As a first step, we simply
overplotted the Chandra positions on the FUV mosaic. Chandra source~7
and FUV source~222 have a close positional match
with a distance of $0.61\arcsec$. This offset
agrees with the $0.4\arcsec$ absolute pointing error of Chandra ($1\sigma$ error) and the additional HST absolute pointing accuracy of $0.1-2\arcsec$.
FUV source~222 is the best CV candidate in
the FUV catalogue, as it is a variable source located between the MS
and WD cooling sequence in the FUV--NUV color-magnitude diagram (CMD), which is the
expected location for CVs \citep{Dieball+05}. This source was expected to show X-ray
emission, and we are confident that Chandra source~7 and FUV
source~222 are the same object. The emission of Chandra source~7 is also consistent with CV emission (see \S\ref{disc}).
We therefore shifted the Chandra positions
so that Chandra source~7 and FUV source~222 match exactly.
No significant rotation of the field of view is expected \cite[see for instance ][]{EGHG03}.
This appears to be the most likely shift in order to align the FUV and X-ray images.

We searched for FUV counterparts within a maximum tolerance radius of $3\sigma$ of the corresponding Chandra source, assuming only detection errors. Note that the full width
half maximum (FWHM) of the FUV PSF is
$< 0.074\arcsec$,
i.e.\ much smaller than the $3\sigma$ error circles for all Chandra
sources (and in most cases smaller than the $1 \sigma$ error circle, see
Table~\ref{tab:chandra_sources}). Therefore we adopted the larger Chandra
error circles as the maximum matching radius.

\begin{figure*}
\centering
\includegraphics[width=10cm]{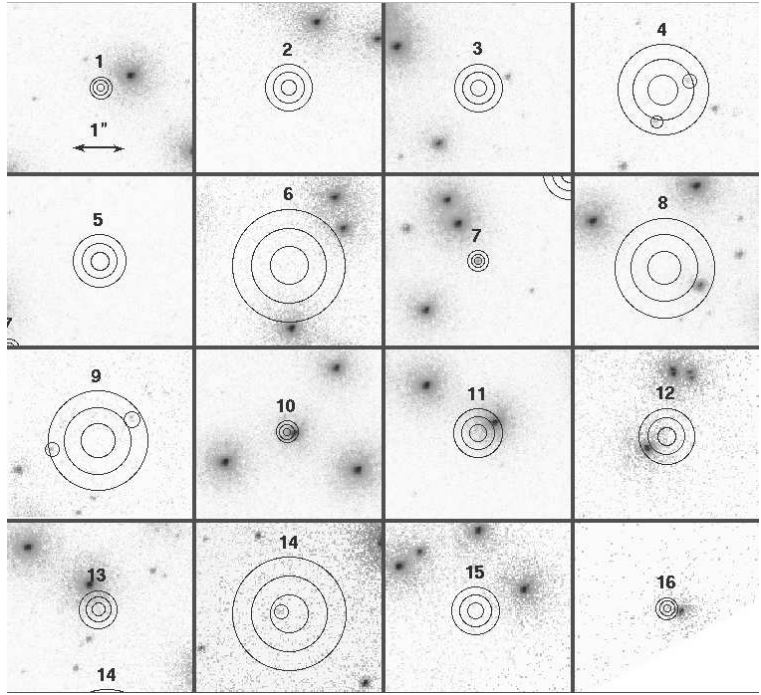}
\caption{Close-ups on the FUV mosaic of the core region of NGC 2808. North is to the top left and East to the bottom left. The overlaid circles and corresponding numbers denote the core Chandra sources with $1\sigma$, $2\sigma$, and
$3\sigma$ error radii. The faint FUV counterpart candidates to Chandra sources~4, 9 and 14 are marked with small circles. The length of $1\arcsec$ is indicated in the top left panel for Chandra source~1. \label{fig:fuvima}}
\end{figure*}

\begin{table*}
\caption{FUV counterpart candidates to the Chandra sources in the core of \mbox{\object{NGC 2808}}. The first column denotes the Chandra source~ID, followed by the FUV source~ID as in \citet{Dieball+05}. The NUV and FUV luminosities are shown in magnitudes and flux density in $10^{-17}$~erg~cm$^{-2}$~s$^{-1}$~\AA$^{-1}$. The approximate ratio between X-ray flux (0.5--8~keV) and FUV and NUV flux densities are given. The distance between the FUV counterpart candidate and the corresponding Chandra source is given in STIS pixels, arcseconds and $\sigma$. Column 11 gives the probability that the specific FUV source is a chance superposition, and the column 12 denotes the probability that, by chance, a FUV source can be found within the $3\sigma$ error circle.}
\label{tab:uvcounterparts}
\centering
\begin{tabular}{ccc@{~~}c@{~~}cc@{~~}c@{~~}cccccc}
\hline\hline                 
Ch. & FUV & \multicolumn{2}{c}{NUV} & $F_X$ & \multicolumn{2}{c}{FUV} & $F_X$ & \multicolumn{2}{c}{Distance} & $\sigma$ & Prob. & Prob. $3\sigma$\\
ID & ID  &     [mag]        &    [flux]      & /$F_{NUV}$ &    [mag]         &     [flux]      & /$F_{FUV}$ &[pixel]&[$\arcsec$]& & $\%$& $\%$\\
\hline
4  & 400 & 19.738$\pm$0.024 &  3.64$\pm$0.08 &   38 & 22.042$\pm$0.118 &   0.55$\pm$0.06 &  253 & 20.43 & 0.505 & 1.74 & 32.65 & 60.53\\
4  & 392 &     --           &    --          &  --  & 21.743$\pm$0.098 &   0.73$\pm$0.06 &  192 & 25.94 & 0.641 & 2.21 & 47.96 &      \\
7  & 222 & 20.824$\pm$0.030 &  1.34$\pm$0.04 & 2168 & 20.906$\pm$0.048 &   1.58$\pm$0.07 & 1839 &  0.00 & 0.000 & 0.00 &    -- &  5.26\\
8  & 111 & 17.428$\pm$0.012 &  3.05$\pm$0.34 &    5 & 18.374$\pm$0.015 &  16.23$\pm$0.22 &    8 & 31.25 & 0.772 & 2.42 & 40.82 & 50.00\\
9  & 476 & 19.583$\pm$0.034 &  4.20$\pm$0.13 &   29 & 21.881$\pm$0.161 &   0.64$\pm$0.10 &  187 & 31.94 & 0.789 & 2.47 & 53.06 & 73.68\\
9  & 457 &     --           &    --          &  --  & 21.646$\pm$0.125 &   0.80$\pm$0.09 &  151 & 37.33 & 0.922 & 2.89 & 62.25 &      \\
10 & 182 & 18.353$\pm$0.009 & 13.03$\pm$0.11 &  200 & 17.039$\pm$0.008 &  55.51$\pm$0.41 &   47 &  6.07 & 0.150 & 2.17 & 5.10 &  2.63\\
11 & 252 & 17.009$\pm$0.004 & 44.92$\pm$0.17 &   13 & 16.400$\pm$0.006 & 100.00$\pm$0.55 &    6 & 16.14 & 0.399 & 2.57 &  6.12 & 18.42\\
12 & 492 & 18.217$\pm$0.016 & 14.76$\pm$0.22 &   29 & 16.783$\pm$0.019 &  70.27$\pm$1.22 &    6 & 17.91 & 0.443 & 2.47 & 20.41 & 34.21\\
14 & 446 & 24.139$\pm$0.668 &  0.06$\pm$0.04 & 1425 & 22.966$\pm$0.226 &   0.24$\pm$0.05 &  381 &  7.08 & 0.175 & 0.48 &  4.08 & 78.95\\
\hline
\end{tabular}
\end{table*}

\begin{table*}
\caption{Optical sources within the $3 \sigma$ error circles of the
 Chandra sources. The first column denotes the Chandra source~ID
 followed by the optical source~ID as in \citet{Piotto+02}, and the FUV~ID. The optical V and B
 magnitudes are given in columns 4 and 6, with the X-ray/optical ratios in columns 5 and 7 (in the same way as in Tab.~\ref{tab:uvcounterparts}), followed by the
 distance in PC pixels (column 8), arcseconds (column 9), and
 $\sigma$ (column 10) between the optical source and the transformed
 Chandra position on the PC image. The final column indicates the
 position of the source in the optical CMD, where BS denotes Blue
 Straggler region, BHB the blue horizontal branch, and MS the faint
 main sequence. \label{tab:opticalcounterparts}} \centering
\begin{tabular}{ccccccccccc}
\hline\hline
Chandra ID &Optical ID& FUV ID & V   & $F_X$/$F_V$    &  B &  $F_X$/$F_B$  &\multicolumn{2}{c}{Distance}&$\sigma$ & CMD \\
           &          &        &[mag]&     & [mag]&  &    [pixel]   & [$\arcsec$] &         &     \\
\hline
3   & 6881    &     -- & 18.445 &  224 & 18.600 &  56 &   2.521   &  0.116      & 0.74   & BS \\
4   & 5315    &    400 & 17.947 &   22 & 18.122 &   6 &  10.949   &  0.504      & 1.72   & BS \\
8   & 6805    &     -- & 15.481 &    2 & 15.491 &   1 &  16.890   &  0.777      & 2.41   & BHB\\
8   & 13062   &     -- & 20.648 &  265 & 20.928 &  74 &  14.739   &  0.678      & 2.10   & MS \\
8   & 7222    &     -- & 20.444 &  220 & 20.686 &  59 &  19.410   &  0.893      & 2.77   & MS \\
9   & 3064    &    476 & 17.729 &   15 & 17.997 &   4 &  17.251   &  0.794      & 2.46   & BS \\
11  & 2486    &    252 & 16.156 &   18 & 16.006 &   3 &   8.763   &  0.403      & 2.57   & BHB\\
11  & 10872   &     -- & 21.260 & 1996 & 21.578 & 579 &   2.789   &  0.128      & 0.82   & MS \\
11  & 2529    &     -- & 20.998 & 1568 & 21.308 & 452 &   3.593   &  0.165      & 1.06   & MS \\
12  & 3165    &     -- & 18.543 &  117 & 18.563 &  25 &   9.697   &  0.446      & 2.47   & BS \\
\hline
\end{tabular}
\end{table*}

\begin{figure}
\centering
\includegraphics[width=\columnwidth]{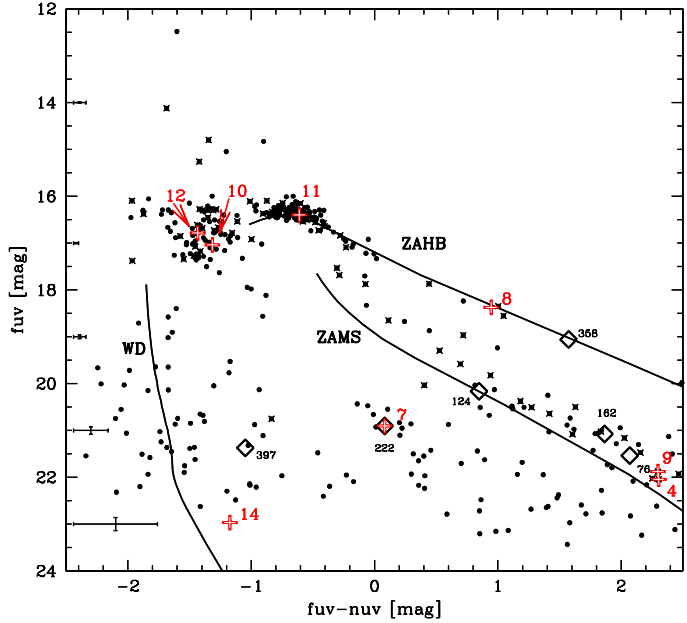}
\caption{FUV--NUV CMD, taken from \citet{Dieball+05}. The variable FUV sources are marked with diamonds and their corresponding FUV ID. Additionally, we marked the FUV sources that constitute possible Chandra counterparts with crosses and numbers. For orientation purposes, we include a theoretical WD cooling sequence, a zero-age main sequence (ZAMS), and a zero-age HB track (ZAHB). Note that the sources located along the ZAMS are BS stars, which are located above the main sequence turn-off and slightly to the red. EHB stars are clustered at the bright end of the ZAHB, and the clump of bright sources slightly to the blue and somewhat fainter are the blue hook stars \citep{Brown+01}. See also \citep{Dieball+05} for an explanation of the theoretical sequences and a discussion of the CMD.\label{fig:fuvnuvcmd}}
\end{figure}

Fig.~\ref{fig:fuvima} shows close-ups on the FUV mosaic.
In total, we found 10 possible FUV
counterparts to 8 X-ray sources, which are listed in
Table~\ref{tab:uvcounterparts}. Their location in the FUV--NUV CMD
is indicated in Fig.~\ref{fig:fuvnuvcmd}.
As can be seen from Fig.~\ref{fig:fuvima}, the $3\sigma$ errors can be
quite large for some Chandra sources. The probability of a false match
between a Chandra and a FUV source correlates with the size of the
Chandra error circle. In order to estimate the probability of such
coincidental matches, we repeatedly shifted the set of Chandra sources
by 90 STIS pixels or $2.2\arcsec$ to the top and left and right. We only allowed for shifts
that kept all Chandra sources within the FUV field of view, so that
the statistics for all Chandra sources are the same.
After each shift, we again checked for FUV counterparts to each
Chandra source. The last two columns in Table~\ref{tab:uvcounterparts} gives the
probability that, just by chance, at least one FUV source can be found
within the same distance as the FUV source that was found
without a random shift, or within the three sigma error circle of that Chandra source.

We used a Monte-Carlo simulation to produce 1\,000 fields with the detected number of UV and X-ray sources, and we deduced the number of matches expected by chance. Using $1\sigma$, $2\sigma$ and $3\sigma$ error circles for the 16 Chandra sources, we found that $0.6\pm0.7$, $2.2\pm1.4$ and $5.2\pm2.4$ matches may be spurious respectively. We thus expect $\sim2-3$ matches to be real among the 10 UV counterparts.
In addition to source 7, two other sources (10 and 14) have probabilities that imply that they may be associated with a FUV source (Table~\ref{tab:uvcounterparts}).

\begin{figure}
\centering
\includegraphics[width=\columnwidth]{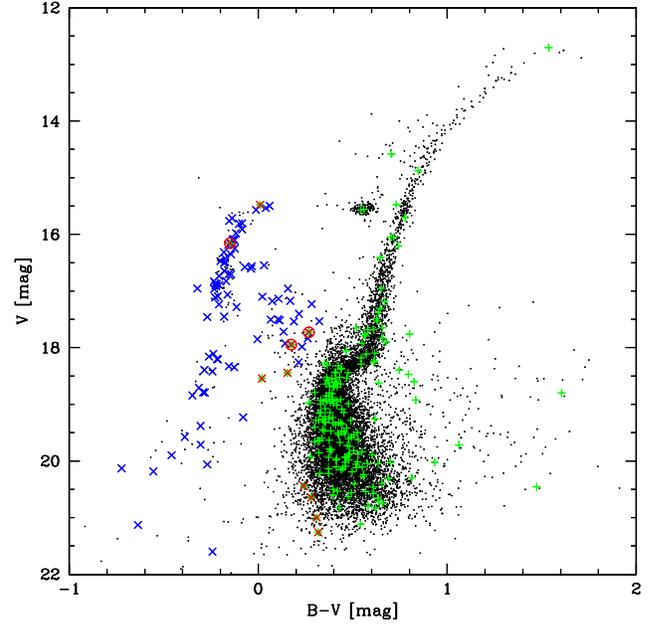}
\caption{Optical CMD. Blue crosses ('x') denote FUV sources seen in optical, green '+' denote optical counterparts to X-ray sources, red crosses ('x') denote the selected optical counterparts listed in Table~\ref{tab:opticalcounterparts}, and the red circled crosses denote the optical counterparts to both FUV and X-ray sources. \label{fig:optcmd}}
\end{figure}

\subsection{Optical counterparts}

\mbox{\object{NGC 2808}} was observed with the HST Wide Field
Planetary Camera~2 (WFPC2) in the F439W and F555W bands, with the
Planetary Camera (PC) centered on the cluster's center. The field of
view of the PC is $36\arcsec\times36\arcsec$.
In order to check for possible optical counterparts to the 16 core
X-ray sources, we used the optical catalogue presented by
\citet{Piotto+02}.

We searched for optical counterparts by transforming all the FUV
coordinates, including the 16 shifted Chandra positions, to the PC
cartesian system, using the \textit{geomap} and \textit{geoxytran}
tasks running under IRAF\footnote{IRAF (Image
 Reduction and Analysis Facility) is distributed by the National
 Astronomy and Optical Observatory, which is operated by AURA, Inc.,
 under cooperative agreement with the National Science Foundation.}
and the FUV--optical matches in the \citet{Dieball+05} catalogue
as a database.
We then looked for optical counterparts within a radius $\le 3\sigma$
of the transformed Chandra errors.
However, nearly all of these are either located on the MS, red giant
branch or red HB. Two of the optical sources are on the BHB and four sources are in
the optical BS region above the MS turn-off. There are four sources
that are faint and blue with $B-V < 0.4$ and $V > 20.4$~mag. An
optical counterpart to a CV is expected to be very blue.
All the above mentioned optical sources are listed in
Table~\ref{tab:opticalcounterparts}
and a colour-magnitude diagram is presented in Fig.~\ref{fig:optcmd}.
Of these, four are optical
counterparts to the FUV sources that might constitute possible matches
to the Chandra sources, see Table~\ref{tab:uvcounterparts}.

\subsection{Ultraviolet counterparts outside the core}

\mbox{\object{NGC 2808}} was observed with the XMM-Newton Optical Monitor (OM) using
the UVM2 filter, centered at 231~nm in the UV band
\citep{Servillat+08-a}. We reprocessed the data in the same way as
\citet{Servillat+08-a}, in order to use the latest version of the SAS (v7.1), and looked for UV
counterparts to the Chandra sources. Because of the crowding of
sources, we could not use this data inside the half-mass radius.

A bright source is seen in the field of view outside the crowded area of
the core of \mbox{\object{NGC 2808}}, which is consistent with the star \mbox{\object{HD 79548}} at
RA$_{2000}$~$9^h11^m33.293^s$,
Dec$_{2000}$~$-$64\degr51\arcmin03.28\arcsec. This star is an A0V star
with magnitudes ${B=10.42}$ and ${V=10.15}$, compatible with the UVM2
magnitude we found. We thus shifted the coordinates of the OM
observation to match this source.
After the shift, the number of possible matches significantly increased from 2 to 16. We used a Monte-Carlo simulation to deduce the number of matches expected by chance, and estimated that $1.6\pm1.2$ matches may be spurious (using $3\sigma$ error circles). This indicates that the nearly all of the matches are real. Moreover, a rotation of the field of view does not seem necessary.

We took into account the $1\sigma$ positional error of the Chandra sources, and the $1\sigma$ positional
error of the UV sources.
The 16 possible counterparts found outside the half-mass radius are listed in Table~\ref{tab:uvom}, nine of which are within the $1\sigma$ error circle. From the nine UV counterparts found
for XMM-Newton sources \citep{Servillat+08-a}, eight are consistent with the
Chandra sources. The XMM-Newton source 29 counterpart is ruled out as the corresponding Chandra source, 33, has a smaller $3\sigma$ error circle which does not match the proposed UV counterpart. Chandra source~36 has a UV counterpart and a matching
XMM-Newton source, but the match between UV and XMM-Newton sources was not
found \citep{Servillat+08-a} because this UV source was not detected with the older version of the SAS.
As they are found outside the half-mass radius of \mbox{\object{NGC 2808}}, these sources are likely to be background or foreground sources (see Sect.~\ref{sec:fore} and \ref{sec:back}).

\begin{table}
\caption{XMM-Newton OM UV counterparts to Chandra X-ray sources. The
  first column corresponds to the Chandra ID and a star indicates that the
  source was detected with XMM-Newton. In column 2 we give the
  distance between the X-ray source~and the possible UV counterpart
  and in column 3 this distance is divided by the $1\sigma$ error. The
  offset from the center of \mbox{\object{NGC 2808}} is given in column 4 and in the
  last column, the UVM2 Vega magnitude.}             
\label{tab:uvom}      
\centering                          
\begin{tabular}{ccccc}        
\hline\hline                 
ID & Distance & $\sigma$ & Offset & UVM2 \\    
\hline                        
  17     &    1.77\arcsec &    2.17 &    0.89\arcmin &   17.00 $\pm$    0.06 \\
  19     &    2.15\arcsec &    2.64 &    1.30\arcmin &   15.18 $\pm$    0.02 \\
  21     &    1.36\arcsec &    1.71 &    1.95\arcmin &   17.25 $\pm$    0.05 \\
  31$^*$ &    0.43\arcsec &    0.54 &    3.19\arcmin &   10.95 $\pm$    0.01 \\
  35     &    0.81\arcsec &    0.94 &    3.36\arcmin &   19.52 $\pm$    0.41 \\
  36$^*$ &    1.34\arcsec &    1.52 &    3.53\arcmin &   19.55 $\pm$    0.40 \\
  39$^*$ &    0.53\arcsec &    0.66 &    3.77\arcmin &   17.83 $\pm$    0.07 \\
  41$^*$ &    0.52\arcsec &    0.65 &    3.86\arcmin &   18.29 $\pm$    0.10 \\
  45     &    2.44\arcsec &    2.83 &    4.02\arcmin &   17.28 $\pm$    0.04 \\
  48$^*$ &    2.39\arcsec &    2.85 &    4.10\arcmin &   17.23 $\pm$    0.04 \\
  49     &    1.26\arcsec &    1.23 &    4.14\arcmin &   19.50 $\pm$    0.43 \\
  51$^*$ &    0.83\arcsec &    0.96 &    4.29\arcmin &   18.99 $\pm$    0.17 \\
  53$^*$ &    0.09\arcsec &    0.12 &    4.42\arcmin &   14.70 $\pm$    0.01 \\
  63     &    0.29\arcsec &    0.33 &    5.15\arcmin &   18.86 $\pm$    0.15 \\
  74$^*$ &    0.11\arcsec &    0.13 &    5.82\arcmin &   17.99 $\pm$    0.07 \\
  99$^*$ &    0.79\arcsec &    0.96 &    8.53\arcmin &   18.29 $\pm$    0.09 \\
\hline                        
\end{tabular}
\end{table}


\section{Discussion}
\label{disc}

In order to identify the X-ray sources linked to \mbox{\object{NGC 2808}}, we first discuss the properties of X-ray sources and their possible counterparts.
The brightest sources with ${L_{[0.5-8keV]}\gtrsim10^{32}\mathrm{~erg~s^{-1}}}$ are likely to be qLMXBs if they are very soft and well fitted by hydrogen atmosphere model, with masses and radii consistent with a neutron star \citep{HGLE03,GBW03b,GBW03}. They are likely to be CVs if they are harder, with $kT>3$~keV \citep{BWO05}, or in this work a photon index less than 2.
X-ray sources with lower luminosities are most likely either CVs, ABs or MSPs, although we note that there are also qLMXBs which have luminosities around ${L_{[0.5-8keV]}\sim5\times10^{31}\mathrm{~erg~s^{-1}}}$ \citep[e.g.][]{HGE05}.
ABs and MSPs are in general expected to be soft \citep{DLSF93,Bogdanov+06}.
The brightest AB observed in a GC reached ${L_{[0.5-8keV]}\sim10^{32}\mathrm{~erg~s^{-1}}}$ \citep{Heinke+05}.
Their spectrum is generally well fitted by a two temperature model with mean $T_{high}\sim10^7$~K and $T_{low}\sim10^6$~K, and they have luminosities of ${L_{X}\sim10^{29}-10^{31}\mathrm{~erg~s^{-1}}}$ \citep{DLSF93}.
MSPs in GCs are well described by a thermal (blackbody or neutron star hydrogen atmosphere) spectrum with a temperature $T_{eff}\sim(1-3)\times10^6$~K, emission radius $R_{eff}\sim0.1-3$~km, and luminosity
${L_{X}\sim10^{30}-10^{31}\mathrm{~erg~s^{-1}}}$ \citep{Bogdanov+06}. Some of them show a significant or dominant non-thermal (power law) component, with spectral photon index $\Gamma\sim1-1.5$, and are more luminous \citep{Bogdanov+06}. The brightest MSP observed in a GC reached ${L_{[0.5-8keV]}\sim1.3\times10^{33}\mathrm{~erg~s^{-1}}}$ \citep[in M28,][]{Becker+03}.

In the UV, CVs are expected to be brighter than qLMXBs, ABs and MSPs, and are located between the MS and the WD cooling sequence in the CMD \citep[e.g.][]{Dieball+05}.
ABs, on the other hand, are expected to be redder than MS stars or on the MS \citep[e.g.][]{EGHG03}.

CVs can show high variability in X-rays on different timescales \citep[e.g.][]{1995cvs..book.....W}.
Large variations of a factor of 5--10, and timescales of hours to days are most commonly associated with dwarf nova outbursts \citep{BWO05,WMM03}. However, such variation can also appear during the quiescent state of some CVs \citep{BWO05}.
ABs show flaring activity. The decay time of a compact flare is $\sim1$~ks, and for two-ribbon flares it reaches 5--20~ks \citep{PPK88}. The rise time is generally shorter than 2~ks \citep{PPK88}. Longer flares have been observed, with a noticeable decay \citep[110~ks,][]{FPT01}, and very exceptional flares can last several days \citep[e.g.][]{KS96}.

\subsection{X-ray sources with a possible UV counterpart in the core}

\subsubsection{Chandra source 7: a CV}

Chandra source~7 has X-ray colour and flux compatible with CV emission
(see Table~\ref{tab:chandra_sources}). Furthermore, its UV counterpart is
located in the gap region between the MS and the WD cooling
sequence in the FUV--NUV CMD (see Fig.~\ref{fig:fuvnuvcmd}).
The X-ray/UV ratios are high for this source (see Table~\ref{tab:uvcounterparts}), similar to those observed for intermediate polars (see Sect.~\ref{sec:XUV_CVs}).

\subsubsection{Chandra source 14: a possible CV}

Chandra source~14 is fainter, with X-ray colour and flux compatible with CV, AB or MSP emission.
Its possible UV counterpart is located in the CV region and the X-ray/UV ratios are compatible with the emission of CVs (see Sect.~\ref{sec:XUV_CVs}).
The probability of finding by chance a FUV source within a $3\sigma$ error circle is high ($\sim80$\%). However, the probability of finding a FUV source just by chance within a $0.5\sigma$ error circle of Chandra source 14 is 4\% (see Table~\ref{tab:uvcounterparts}), therefore it seems likely that FUV source 446 is the true counterpart, making Chandra source 14 a good CV candidate. From simulations, the probability of having a FUV source which is also one of the $\sim60$ CV candidates is 0.5\% .

\subsubsection{Chandra source 10: another CV?}

Chandra source~10 has very similar X-ray parameters to Chandra source~7,
a priori suggesting that this source also constitutes a CV (Fig.~\ref{fig:chandra_cmd}).
Its possible UV counterpart is bright in FUV and NUV,
and it is located in the faint part of the EHB region in the FUV--NUV CMD.
Although field CVs in outburst can be as bright as this source in FUV \citep[see Fig.~3 in][]{Dieball+05}, in this case the source would have to be in outburst for $\sim30$~days, or during each FUV observation \citep[2000 January 18 and 19, and February 16 and 20, ][]{Dieball+05}, which is unlikely as CV outbursts are expected to be rare in GCs \citep{Shara+96}.
It is also possible that its true UV counterpart is masked by the
bright, close UV source.

\subsubsection{Chandra source 11}

The flux and colour of Chandra source~11 are consistent with a CV or an AB
candidate.
It possibly matches with a bright UV source, located in the BHB
clump in the FUV--NUV CMD, and an optical source which also suggests
that this source is a BHB star (Table~\ref{tab:uvcounterparts}). X-ray emission is not expected from
such HB stars, and this source could constitute another kind of exotic
binary. The matching distance between the X-ray and FUV
coordinates is $2.6\sigma$. It is possible that the UV
source is simply a mismatch.
One of the optical counterparts (\#10872,
Table~\ref{tab:opticalcounterparts}) is blue, as expected for a CV.

\subsubsection{Chandra source 12}

Chandra source~12 has a similar flux and colour to Chandra source~11,
consistent with the CV or AB hypothesis. Its luminosity also possibly varied between XMM-Newton and Chandra observations ($2\sigma$).
The possible UV counterpart is located in the EHB clump in the FUV--NUV CMD.
However, the match to this FUV source is within
$2.5\sigma$, with a probability of $\sim18\%$ that this match is a
chance superposition. This might well be a mismatch and the true FUV
counterpart was not detected in our FUV images.
One of the possible optical counterparts is in the BS region (see Table~\ref{tab:opticalcounterparts} and Fig.~\ref{fig:optcmd}).

\subsubsection{Chandra source 8}

Chandra source 8 can be either a CV, an AB, or a MSP candidates.
It possibly matches with a bright FUV source,
located on the HB sequence in the FUV--NUV CMD. This source is
thus not expected to reach such a luminosity in X-rays and it could be a mismatch.
We found two possible optical counterparts compatible with the expected emission of a CV (see Table~\ref{tab:opticalcounterparts} and Fig.~\ref{fig:optcmd}).

\subsubsection{Chandra sources 4 and 9}

Chandra sources 4 and 9 are fainter X-ray sources, giving weak
constraints about their nature. They are either CV, AB,
or MSP candidates.
Both Chandra sources 4 and 9 have two possible FUV counterparts. Of
these, FUV sources 400 and 476 seem to be the likely counterparts
to Chandra source 4 and 9, respectively, as they are closer
(Table~\ref{tab:uvcounterparts}).
These FUV sources are located at the faint end of the BS sequence. The
discrimination between BS and CV candidates is difficult, and both
populations might well overlap in the FUV--NUV CMD. Thus, it is
possible that X-ray sources 4 and 9 are CVs, or they might be
X-ray emitting BSs. Such an object was found in 47~Tuc by
\citet{Knigge+06}. They suggested that this exotic object
is a detached binary consisting of a BS primary with an X-ray-active
MS companion whose formation would involve at least three stars.
\citet{EGHG03} and \citet{Heinke+05} also identified such objects in 47~Tuc.
The coincidence probability of these sources is high (Table~\ref{tab:uvcounterparts}) and they could be mismatches.

\subsection{X-ray sources without a UV counterpart in the core}

\subsubsection{Chandra source 1: a qLMXB candidate}

Chandra source~1 corresponds to XMM-Newton source~C2, which was presented as a qLMXB candidate by \citet{Servillat+08-a}. The Chandra flux and colour of this source also support the idea that it is a qLMXB. Moreover, the spectrum of Chandra source~1 is well fitted with a hydrogen atmosphere model with parameters consistent with a neutron star. No hard tail is detected in the spectrum, as in general in GC qLMXBs \citep{Heinke+03a}.
The upper limit is ${F_{[0.5-8keV]}\sim5\times10^{-16}\mathrm{~erg~cm^{-2}~s^{-1}}}$ ($<2.5$\% of the total flux), assuming a power law model with photon index 1--1.5 \citep{Campana+98}, and 1 photon detected in the band 2--8~keV.
\citet{Servillat+08-a} estimated that $3\pm1$~qLMXBs are expected in the core of \mbox{\object{NGC 2808}}, using the empirical correlation between the number of qLMXBs and the star encounter rate. We found no evidence in our Chandra observation for other qLMXB candidates.

\subsubsection{Chandra source 16: a variable CV}

Chandra source 16 is a variable X-ray source. This variability is observed on different time scales: during the first Chandra observation (few hours), between the two Chandra observations (1~day), and between XMM-Newton and Chandra observations (28~months). Its flux, colour and variability in X-rays suggest it could be a CV, possibly showing the signature of a CV outburst in X-rays \citep[e.g.][]{BWO05}.
An AB interpretation is less probable as even in long flares, the decay of the flare would be detectable over the first 45~ks of Chandra observation \cite[as in][for a 110~ks flare]{FPT01}.

\subsubsection{Chandra source 3: possible CV}

Chandra source~3 is very hard and shows short-term variability. The lightcurve indicates a count increase by a factor $\sim5$ between the beginning of the first Chandra observation and the second Chandra observation (Fig.~\ref{fig:chandra_lcvar}). The rise time is longer than 10~ks (end of first Chandra observation, Fig.~\ref{fig:chandra_lcvar}), exceptional for an AB flare \citep[e.g.][]{PPK88}. This kind of fluctuation would be more typical of a CV, and it is possibly the signature of a CV outburst in X-rays \citep[e.g.][]{BWO05}.
We note that a possible optical counterpart falls in the BS region in Fig.~\ref{fig:optcmd}, however it could be a mismatch.

\subsubsection{XMM-Newton source C5}

XMM-Newton source~C5, which was not detected during our Chandra observation, was hard, and is variable over 28 months. The flux varied by a factor of at least~5 over this period.
From its flux and colour, it could be a variable CV or an AB which showed a flare. No variability was detected from this source during the XMM-Newton observation of $\sim30$~ks with the \textit{pn} detector \citep{Servillat+08-a}. Such a long flare would be exceptional for an AB, although it cannot be excluded \citep[e.g.][]{KS96}. CVs routinely show fluctuations in X-rays, some of them being associated with CV outbursts \citep[e.g.][]{BWO05}.

\subsubsection{Chandra source 17}

We note that Chandra source~17, which is close to the half-mass radius
of \mbox{\object{NGC 2808}}, has colours and fluxes consistent with those of a CV, an AB or a MSP.
It shows variability that is possibly consistent with a CV fluctuation or an AB flare, as it lasted $\sim10$~ks (Fig.~\ref{fig:chandra_lcvar}).
This source could therefore be linked to the cluster, and could be a CV or an AB.
If it is a CV, its offset from the core would make it a primordial CV candidate \citep{HAS07}.

\subsubsection{Other Chandra core sources}

From their luminosities and colours, Chandra sources 2, 5, 13, 15 are consistent with the CV or AB hypothesis. Similar X-ray sources are identified as CVs or ABs by \citet{Heinke+05} in \mbox{\object{47 Tuc}}. However, we did not find FUV counterparts, nor detected variability, and we cannot confirm their nature.
Chandra source~6 is fainter, and is consistent with being a CV, an AB, or even a bright MSP candidate.

\subsection{Variable sources in the core}

Several authors have claimed that there is a lack of CV outbursts in GCs \citep[e.g. ][]{Shara+96,EGHG03b,DLM06}.
To explain this deficit, it was proposed that most GC CVs are moderately magnetic with low accretion rates, systems for which very few outbursts are expected \citep{Grindlay99,Ivanova+06,DLM06}.

\citet{Shara+96} looked for dwarf nova outbursts in 47~Tuc using HST observations for a cumulative time of $\sim130$~ks. They found only one outburst and showed that they should have detected at least 12 from simulations, implying a lack of outbursts from GC CVs.
Outbursts of GC CVs have also been observed in UV for two magnetic CVs in \mbox{\object{NGC 6397}} \citep{Shara+05}. The cumulative time of their observation was $\sim66$~ks. Our X-ray observations lasted $\sim90~ks$ (XMM-Newton and Chandra), and could therefore lead to the detection of dwarf nova outburst signatures.

An optical outburst of a dwarf nova can be linked to an increase of X-ray emission (factor $\sim5$), followed by a drop (factor $\sim3$) due to the variation of thickness of the boundary layer at different accretion rates \citep{WMM03,1995cvs..book.....W}. However, X-ray fluctuations of CVs are also detected during the quiescent state \citep{BWO05}.

Our observation indicates that some sources in the core show such fluctuations in X-rays.
Chandra sources~3, 16 and XMM-Newton source C5 are likely to be CVs that possibly showed signatures of outbursts in X-rays.
Simultaneous X-ray and UV (or optical) observations should be performed for several GCs to confirm these possible dwarf nova outbursts and assess the rate of outbursts in GCs.

\subsection{Expected X-ray sources linked to \mbox{\object{NGC 2808}}}
\label{sec:expected}

Assuming a completeness in the detection of sources at a luminosity of ${L_{[0.5-8keV]}\sim2\times10^{31}\mathrm{~erg~s^{-1}}}$, we can estimate the number of expected X-ray sources in \mbox{\object{NGC 2808}}.

To estimate the number of expected X-ray sources in \mbox{\object{NGC 2808}}, we use a general approach based on the correlations given by \citet{PH06}, where they used the results of observations of 22~GCs with Chandra. Taking into account the specific encounter frequency of \mbox{\object{NGC 2808}}, as defined in \citet{PH06}, and our completeness limit, we expect $30\pm6$ X-ray sources in \mbox{\object{NGC 2808}}. The error is extrapolated from the $1\sigma$ error in \citet{PH06}.
Of these sources, $3\pm1$~qLMXBs are expected, and $17\pm3$~CVs with a luminosity greater than ${L_{[0.5-8keV]}\sim4.25\times10^{31}\mathrm{~erg~s^{-1}}}$ \citep[corresponding to populations I and II respectively, as defined in][]{PH06}.
However, we detected only 11 sources in the half-mass radius of \mbox{\object{NGC 2808}} and above the completeness limit.

Looking only at CVs, we can compare our results to the population synthesis simulation performed by \citet{Ivanova+06}. They estimated the number of CVs and detectable CVs expected in a GC similar to \mbox{\object{NGC 2808}} (their standard model): same core density, relaxation time and metallicity, but velocity dispersion and escape velocity somewhat lower in their model.
Their simulation lead to 209 CVs formed after 10~Gyr,
of which 47 have a luminosity higher than ${L_{[0.5-8keV]}\sim3.6\times10^{30}\mathrm{~erg~s^{-1}}}$.
The number of detectable CVs seems stable over an interval of ages of 7--14~Gyr with a standard deviation of $4$.
We can adapt this result to our completeness limit by applying a factor $\sim0.36$ \citep[obtained from the number of sources detected in \mbox{\object{47 Tuc}} with the corresponding limiting luminosities in][]{Heinke+05}.
This leads to $17\pm4$ detectable CVs predicted for our observation. We detected 8 (and possibly up to 15) CV candidates, which seems less than the predicted number of CVs from simulations.
In the same way, \citet{Knigge+08} found that there are only a few CVs among the bright
FUV sources located between the MS and WD cooling sequence in
\mbox{\object{47 Tuc}}, indicating a possible deficit of CVs compared to what is expected.
If this is true, our understanding of CV formation and evolution in GCs might have to be revised.

We compared our results to the observation of \mbox{\object{47 Tuc}} \citep{Heinke+05}. \mbox{\object{47 Tuc}} is similar in mass, density and concentration to \mbox{\object{NGC 2808}} \citep{Harris96}.
The specific encounter frequency of \mbox{\object{NGC 2808}} is comparable to that of \mbox{\object{47 Tuc}}.
\mbox{\object{NGC 2808}} has an intermediate metallicity \citep[\mbox{$\rm{[Fe/H]} = -1.36$} corresponding to ${Z\sim0.001}$, ][Table~2]{Yi+01}, while \mbox{\object{47 Tuc}} has a higher metallicity ($\rm{[Fe/H]} = -0.76$, or ${Z\sim0.0035}$).
About $31\pm3$~X-ray sources above our completeness limit \citep[the error corresponds to the $1\sigma$ error on the luminosity in][]{Heinke+05} were detected in the half-mass radius of \mbox{\object{47 Tuc}}, of which: 2--5~qLMXBs, 16--19~CVs, 4--5~ABs, and 1~MSP.

\mbox{\object{M80}}, has been observed with Chandra \citep{Heinke+03}. Its metallicity is lower ($\rm{[Fe/H]} = -1.75$, or ${Z\sim0.0004}$), and its mass, density and concentration are close to \mbox{\object{NGC 2808}} values \citep{Harris96}. A total of $17\pm2$ sources were detected above a similar luminosity threshold to our observation, of which 2 are qLMXB candidates, and 5 possible CVs. This is similar to our number of detections in \mbox{\object{NGC 2808}}.

The possible deficit of X-ray sources could therefore be linked to specific parameters of \mbox{\object{NGC 2808}}.
Metallicity seems to be a key parameter that could highly affect the number of X-ray sources in GCs at a given age. Due to the lower opacity, metal poor stars are generally hotter and more compact.
This could lead to the formation of WDs with different properties, and will also determine if, when and how mass transfer occurs in binaries \citep{demink+07}.
\mbox{\object{NGC 2808}}, \mbox{\object{47 Tuc}} and \mbox{\object{M80}} have very different metallicities, which could explain the differences in the number of detectable X-ray sources.
In the same way, \citet{KMZ07} found that metal-rich extragalactic GCs host three times as many LMXBs than metal-poor ones.

\mbox{\object{NGC 2808}} presents unusual features in the optical: an extended BHB
with clumps \citep{Bedin+00}, several MS corresponding to different
star populations \citep{Piotto+07}, and abundances anomalies in HB
stars \citep{Pace+06}. All these particular features are linked in
some way to the metallicity of the stars. \citet{Piotto+07} proposed
that several populations of stars were formed successively, increasing
the helium and metal content of the cluster material at each
round. The possible deficit of X-ray sources we observed is possibly linked to this
specific evolution of \mbox{\object{NGC 2808}} which modified its metallicity content.

\subsection{X-ray and UV emission from CVs}
\label{sec:XUV_CVs}

\begin{figure}
\centering
\includegraphics[width=\columnwidth]{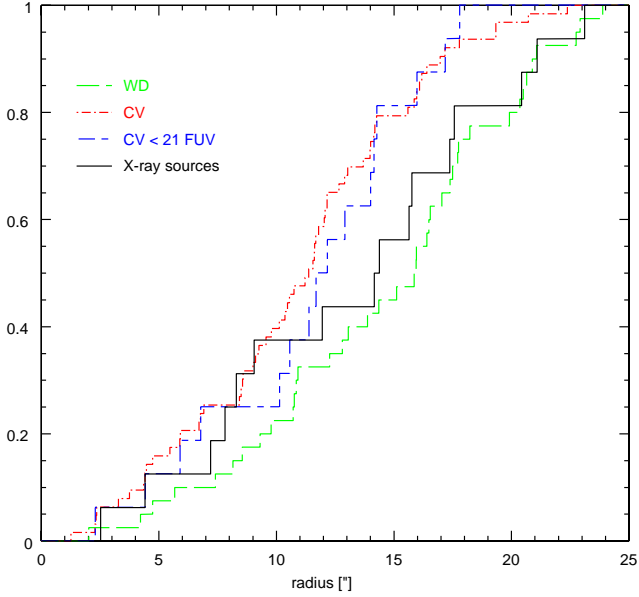}
\caption{Cumulative radial profiles of different populations in \mbox{\object{NGC 2808}}. The populations of WDs and CVs correspond to FUV selected sources \citep{Dieball+05}. \label{fig:radial_profiles}}
\end{figure}

\citet{Knigge+08} showed that no more than half of the objects lying in the CV region in the FUV-NUV diagram of \mbox{\object{47 Tuc}} are confirmed as CVs.
If the types of object in the gap in \mbox{\object{NGC 2808}} have similar proportions as in \mbox{\object{47 Tuc}}, then we expect that $\sim30$ of the $\sim60$ CV candidates detected \citep{Dieball+05} will ultimately turn out to be CVs.
With Chandra, we obtained at most 15 CV candidates.
If we take into account only significant
matches, we found two UV counterparts that have UV properties clearly compatible
with the CV hypothesis (Chandra sources 7 and 14).

Even if we take into account the incompleteness of our observations, it seems that X-ray and UV emission from CVs are decorrelated, as the brightest X-ray sources in \mbox{\object{NGC 2808}} are generally not the brightest FUV sources.
The four CVs confirmed in FUV by \citet{Knigge+02} in \mbox{\object{47 Tuc}}
also have very different X-ray to FUV ratios, strengthening this idea. For instance, AKO~9 is brighter than V1 in FUV \citep[magnitudes $\sim16$ and $\sim18$ respectively,][]{Knigge+02}, but fainter in X-rays \citep[${L_{\mathrm{0.5-6~keV}}\sim5\times10^{31}}$ and ${6.8\times10^{32}\mathrm{~erg~s^{-1}}}$ respectively,][]{Heinke+05}.

The cumulative radial profiles of the X-ray population and
the FUV CV candidates are shown in Fig.~\ref{fig:radial_profiles}.
The spatial distributions do not appear to be significantly different with a KS test. As there is a mass segregation effect in the cluster, this would indicate that the mean mass of the systems is not a dominant parameter to explain the different properties of these populations.

UV and X-ray photons come from different processes.
The UV emission is mainly due to the accretion disk \citep[non-magnetic CVs,][]{Godon+08}, accretion curtains (intermediate polars) or accretion streams (polars), and possibly the hot spot. In systems with low accretion rates or in magnetic CVs, the WD also contributes to the UV emission \citep{MS84}.
On the other hand, the X-ray emission arises from close to the surface of the WD \citep{WW03}, produced by the boundary layer \citep{Pringle77,PR85} or the shock above the WD magnetic pole.

Therefore long period non-magnetic nova-like CVs (large accretion disk with high accretion rate) tend to be relatively X-ray faint and UV bright.
Non-magnetic systems should generally be bright in UV due to their dominant accretion disc.
ROSAT observations of field CVs in X-ray and optical strengthen this idea \citep{Verbunt+97}.
Contrary to this, magnetic CVs seem to be more luminous in X-rays than in UV \citep[e.g.][]{Verbunt+97,EGHG03b}. Part of the UV emission is expected to be suppressed in these systems due to the truncation of the inner portion of the accretion disc of intermediate polars \citep{Grindlay99}. This is also observed for polars which have no accretion disks \citep[e.g.][]{1995cvs..book.....W}.

FUV source~397 is variable in FUV, as is FUV source~222, and is a CV candidate. However, no counterpart is detected in X-ray, contrary to FUV source~222 which matches with Chandra source~7. These CV candidates could therefore be in separate CV classes. FUV source~397 is likely to possess a UV-bright accretion disk as expected for non-magnetic systems. Chandra source~7 is brighter in X-ray as expected for magnetic systems, and variable and bright in UV due to the probable presence of an accretion disk. It is therefore likely to be an intermediate polar.

We estimated a $F_X$/$F_{NUV}$ ratio for several CVs belonging to different classes, as estimated in Table~\ref{tab:uvcounterparts}, where $F_{NUV}$ is the flux density between 2500--3000~\AA, and $F_X$ the flux in the band 0.5--8~keV. Polars have ratios greater than 5000 \citep[ratios extrapolated from][]{RC03,Ramsay+04,Vogel+08}. Intermediate polars appear to have a ratio greater than 2000 \citep{HMZ02,deMartino+05,deMartino+06}.
The detection limit of the NUV observation is ${6\times10^{-19}\mathrm{~erg~cm^{-2}~s^{-1}~\AA^{-1}}}$, and the limit in X-rays is ${9\times10^{-16}\mathrm{~erg~cm^{-2}~s^{-1}}}$. Therefore, the X-ray/NUV ratio for the CV candidates detected in UV is lower than $\sim1500$. The $\sim30$ CV candidates detected in UV and not in X-rays are thus likely to be mostly non-magnetic systems \citep[such as the dwarf nova YZ~Cnc with a ratio of $\sim500$,][]{Hakala+04}.
Most intermediate polars in the field are more luminous than ${10^{31}\mathrm{~erg~cm^{-2}~s^{-1}}}$ in X-rays \citep[][see also the Intermediate Polar Home Page\footnote{http://asd.gsfc.nasa.gov/Koji.Mukai/iphome/iphome.html} maintained by K. Mukai, where 12 out of 14 have luminosities above this limit]{Verbunt+97}. We should have detected most of these in our Chandra observation if their emission is similar to intermediate polars in the field. This would lead to a maximum of $\sim14$ intermediate polars (we exclude Chandra source 14 whose X-ray/NUV ratio is lower than 2000, and Chandra source 1 which is a qLMXB candidate). The proportion derived is $\sim30\%$ of the detected CV candidates, and $\sim7\%$ of the expected GC CV population, estimated to be $\sim200$~CVs \citep[][see also \S\ref{sec:expected}]{Ivanova+06}.
This is coherent with the proportion of intermediate polar candidates in the field, which can be estimated to $\sim5\%$ from the catalogue of \citet[][updated Feb. 2008]{RK03}.
Due to the incompleteness of our observations, this result does not allow us to confirm or rule out a possible excess of intermediate polars in \mbox{\object{NGC 2808}}. However, with a deeper sample, this method could allow us to better quantify the proportion of intermediate polars.

\begin{figure*}
\centering
\hbox{
\includegraphics[width=\columnwidth]{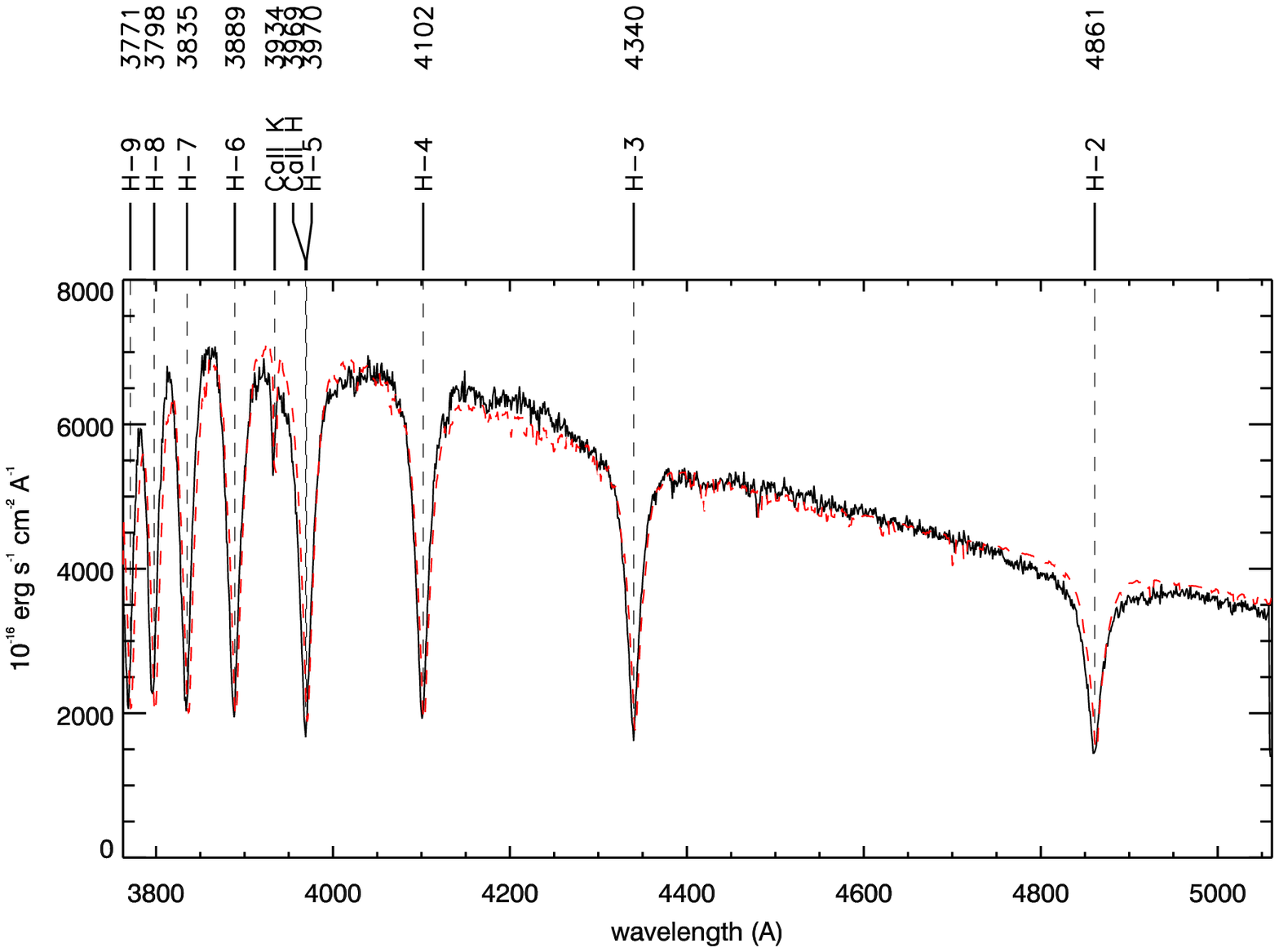}
\includegraphics[width=\columnwidth]{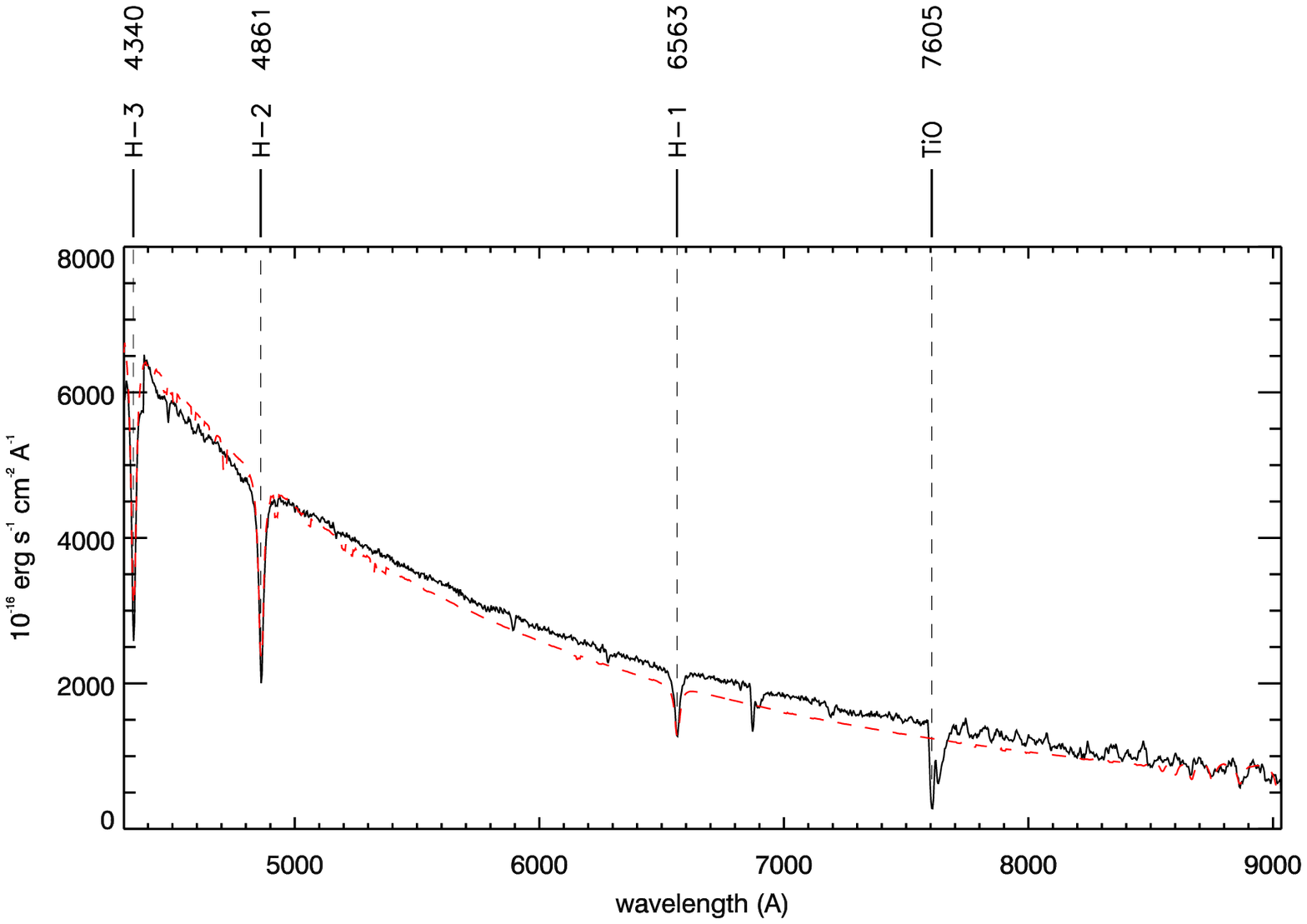}
}
\caption{Dereddened spectra of \mbox{\object{HD 79548}}. The model of an A0V star is shown in dashed line \citep[updated July 2001]{1993yCat.6039....0K} .\label{fig:hd79548}}
\end{figure*}

\subsection{Constraint on the mass of an intermediate mass black-hole in \mbox{\object{NGC 2808}}}

Following \citet{Servillat+08-a}, we can constrain the possible mass of an intermediate mass black-hole (IMBH) in \mbox{\object{NGC 2808}}. No sources are detected at the center of mass of \mbox{\object{NGC 2808}}, where such a massive object is expected to be found.
We assumed that the BH is fed by intracluster gas with a density of $\sim0.5\mathrm{~cm^{-3}}$ derived from \citet{PR01} with \object{NGC~2808} parameters. We considered that the IMBH is radiatively inefficient as for an optically thin advection-dominated accretion flow \citep[ADAF, see e.g.][]{NMQ98}. Following \citet{GHEM01}, our limiting luminosity implies an upper limit of $\sim140\mathrm{~M_{\sun}}$ for a central IMBH in \object{NGC~2808}.
\citet{MS08} used ATCA radio observations to put a mass limit on an IMBH in \mbox{\object{NGC 2808}} (370~M$_{\sun}$). They cast doubt on suggestions that globular clusters may follow the same $M_{BH}-\sigma$ relation as galaxies.

\subsection{Foreground sources}
\label{sec:fore}

\subsubsection{Chandra source 31}

Chandra source 31 has a bright UV and optical counterpart compatible
with the A0V star \mbox{\object{HD 79548}}.
The distance estimate of \mbox{$\sim800$~pc} to this source \citep{Servillat+08-a} leads to an X-ray
luminosity of
${L_{\mathrm{0.5-8~keV}}=7.3\pm1.9\times10^{29}\mathrm{~erg~s^{-1}}}$. The
X-ray emission is soft and showed some variation since the XMM-Newton observation.The luminosity is in agreement with the observed luminosity of X-ray bright A-type stars \citep{SS07}.
However, the source of the emission may come from a late-type active star companion, as the majority of A-type stars are expected to be particularly X-ray dark \citep{SS07}.
Using VLT/FORS1 spectra (see Fig.~\ref{fig:hd79548}), we note that a contribution from a late-type star is possible, and difficult to detect as its flux could be at least 100 times lower than the A-star.
The 12 radial velocity measurements of the H$\beta$ line do not show evidence for a period, so we cannot confirm the presence of a companion.

\subsubsection{Chandra source 113}

Chandra source 113 showed a flare in its X-ray lightcurve (Fig.~\ref{fig:chandra_lcvar}), which reached at least three times the mean flux of the source. The decay timescale was $\sim5$~ks.
The photon index of the spectrum indicates a soft source (2.29, Table~\ref{tab:chandra_sources}). These characteristics are in agreement with those of a chromospherically active star \citep{BP03}.

\subsection{Background sources}
\label{sec:back}

\subsubsection{Sources with radio counterparts}

\mbox{\object{NGC 2808}} was observed with the Australia Telescope Compact Array (ATCA) on 24 January 1992. The data processing is described in \citet{MS08}.
Chandra source~73 has a radio counterpart. It is the most luminous X-ray source of the field, its X-ray emission is hard and absorbed, and it showed a colour variation between XMM-Newton and Chandra observations.
Chandra sources~50 and 89 also have a radio counterpart, and possibly Chandra source~107 and XMM-Newton source~8 (outside the field of view of Chandra). All these sources are hard and absorbed in X-ray (Table~\ref{tab:chandra_sources}), as expected for AGN \citep{Mainieri+07}.

\subsubsection{Chandra source 99}

This source varied in luminosity between the XMM-Newton and Chandra observations. It has been proposed to be a background AGN based on its hard, absorbed X-ray spectrum \citep{Servillat+08-a}. The Chandra observation is consistent with this hypothesis, as is the UV emission detected with the XMM-Newton OM. We detect an infrared counterpart in the Spitzer data found in the archives\footnote{http://ssc.spitzer.caltech.edu/archanaly/archive.html} (AOR 11586048), which is not compatible with the black body emission of a star. These properties are consistent with the emission of a galaxy, possibly an ultraluminous infrared galaxy \citep{LFS06,Braito+04}.


\section{Conclusions}

We presented Chandra observations of \mbox{\object{NGC 2808}} coupled with previous
XMM-Newton observations \citep{Servillat+08-a}, HST FUV
observations \citep{Dieball+05}, VLT and ATCA observations.
We have shown that 16 Chandra sources are likely to be linked to the
cluster, with possibly a 17th close to the half mass radius.
One of these is consistent with the X-ray emission of a
qLMXB, confirming the previous detection with XMM-Newton. Two Chandra
sources (7 and 14) have FUV counterparts that show emission compatible with a CV.
Chandra source 10 is likely to be a CV from its X-ray emission, but no UV counterparts was found to confirm its nature.
Another highly variable source (16) in the core is
likely to be a CV, as well as two other variable sources (Chandra 3 and XMM-Newton C5).
Two other Chandra sources (8 and 11) have optical counterparts compatible with the expected emission of CVs.
We have thus identified 7 CV candidates (plus XMM-Newton source C5) and the observations indicate that there may be
as many as 15 in the Chandra observations (although some of the faintest may
be ABs or MSPs), along with $\sim30$ CV candidates in the HST UV observations.
This significant population of close binaries is likely to play an important role in slowing down the core collapse of this cluster.
Compared to the number of X-ray sources detected in \mbox{\object{47 Tuc}} and expected from dynamical formation, we found a possible deficit of X-ray sources in \mbox{\object{NGC 2808}}.
This might indicate a true deficit of CVs, which is possibly linked to the
metallicity content and the complexity of the evolution of \mbox{\object{NGC 2808}}.


\begin{acknowledgements}
This paper has strongly benefitted from the careful review by the referee, C. Heinke, for which we are very grateful.
MS is grateful to the University of Southampton and the Astronomy
group for hosting him for two months, where part of this work was done.
MS thanks T. J. Maccarone for interesting discussions and for the processing of ATCA data which improved the content of this article.
This research has made use of data obtained from the Chandra Data Archive and software provided by the Chandra X-ray Center.
This work is also based on observations obtained with XMM-Newton, an ESA science mission with instruments and contributions directly funded by ESA Member States and NASA.
We thank the CNES for support of the operational phase of this mission.
Part of this work is based on: FORS1 observations collected with the Very Large Telescope at the European Southern Observatory, Cerro Paranal, Chile; observations with the Australia Telescope Compact Array, founded by the Commonwealth of Australia for operation as a National Facility managed by the CSIRO; and observations made with the Spitzer Space Telescope, which is operated by the Jet Propulsion Laboratory, California Institute of Technology under a contract with NASA.
\end{acknowledgements}






\longtabL{1}{
\begin{landscape}
\begin{longtable}{cccccccccccccc}
\caption{\label{tab:chandra_sources}Detected Chandra sources in the direction of \mbox{\object{NGC 2808}}. The columns correspond to the Chandra ID, the detection method (wavdetect or by hand), the XMM-Newton ID, position in RA and Dec (2000), the $1\sigma$ detection error (Error), the number of Counts (Cts), the Chandra and XMM-Newton fluxes in ${\times10^{-14}\mathrm{~erg~cm^{-2}~s^{-1}}}$, the Chandra and XMM-Newton hardness ratios (HR) using the bands 0.5--2 and 2--8~keV. The best fit using an absorption (N$_H$) and a power law of photon index (PH) is also given, and the last column gives the KS probability of constancy.}\\
\hline\hline
ID & Det. & XMM & RA$_{2000}$ & Dec$_{2000}$ & Error & Cts & Flux & XMM flux & HR & XMM HR & N$_H$ & PH & KS \\
\hline
\endfirsthead
\caption{continued.}\\
\hline\hline
ID & Det. & XMM & RA$_{2000}$ & Dec$_{2000}$ & Error & Cts & Flux & XMM flux & HR & XMM HR & N$_H$ & PH & KS \\
\hline
\endhead
\hline
\endfoot
    1 & w & C2 &   9$^h$12$^m$01.91$^s$ & $-$64$\degr$51$\arcmin$50.67$\arcsec$ &  0.05 &    98 &      2.23 $\pm$     0.47 &      3.60 $\pm$     0.54 &  $-$0.96 $\pm$  0.03 &  $-$0.68 $\pm$  0.05 &   0.12 &   3.74 &  6.25E-02 \\
    2 & w & -- &   9$^h$12$^m$01.72$^s$ & $-$64$\degr$51$\arcmin$41.25$\arcsec$ &  0.11 &    22 &      0.55 $\pm$     0.32 &       --  &  $-$0.27 $\pm$  0.21 &     --   &   0.12 &   1.47 &  4.11E-01 \\
    3 & w & -- &   9$^h$12$^m$03.32$^s$ & $-$64$\degr$51$\arcmin$40.08$\arcsec$ &  0.12 &    20 &      0.90 $\pm$     0.46 &       --  &   0.50 $\pm$  0.19 &     --   &   0.12 &   0.18 &  2.62E-03 \\
    4 & w & C2 &   9$^h$12$^m$01.28$^s$ & $-$64$\degr$51$\arcmin$49.63$\arcsec$ &  0.22 &     6 &      0.14 $\pm$     0.12 &      3.60 $\pm$     0.54 &  $-$0.33 $\pm$  0.38 &  $-$0.68 $\pm$  0.05 &   0.12 &   1.74 &  2.08E-01 \\
    5 & w & C1 &   9$^h$12$^m$04.02$^s$ & $-$64$\degr$51$\arcmin$49.01$\arcsec$ &  0.13 &    17 &      0.53 $\pm$     0.43 &     10.42 $\pm$     0.83 &   0.06 $\pm$  0.24 &  $-$0.13 $\pm$  0.04 &   0.12 &   0.90 &  8.64E-01 \\
    6 & h & -- &   9$^h$12$^m$01.29$^s$ & $-$64$\degr$51$\arcmin$39.52$\arcsec$ &  0.28 &     4 &      0.09 $\pm$     0.08 &       --  &  $-$0.50 $\pm$  0.43 &     --   &   0.12 &   1.89 &  3.20E-01 \\
    7 & w & C1 &   9$^h$12$^m$04.39$^s$ & $-$64$\degr$51$\arcmin$48.23$\arcsec$ &  0.05 &   109 &      2.90 $\pm$     0.81 &     10.42 $\pm$     0.83 &  $-$0.14 $\pm$  0.09 &  $-$0.13 $\pm$  0.04 &   0.12 &   1.35 &  9.67E-01 \\
    8 & h & -- &   9$^h$12$^m$03.86$^s$ & $-$64$\degr$51$\arcmin$38.50$\arcsec$ &  0.25 &     5 &      0.14 $\pm$     0.13 &       --  &  $-$0.20 $\pm$  0.44 &     --   &   0.12 &   1.15 &  5.57E-01 \\
    9 & h & -- &   9$^h$12$^m$01.40$^s$ & $-$64$\degr$51$\arcmin$56.97$\arcsec$ &  0.25 &     5 &      0.12 $\pm$     0.11 &       --  &  $-$0.20 $\pm$  0.44 &     --   &   0.12 &   1.57 &  7.95E-01 \\
   10 & w & C1 &   9$^h$12$^m$04.84$^s$ & $-$64$\degr$51$\arcmin$47.50$\arcsec$ &  0.05 &    99 &      2.60 $\pm$     0.90 &     10.42 $\pm$     0.83 &  $-$0.21 $\pm$  0.10 &  $-$0.13 $\pm$  0.04 &   0.12 &   1.38 &  9.78E-01 \\
   11 & w & C1 &   9$^h$12$^m$04.83$^s$ & $-$64$\degr$51$\arcmin$51.69$\arcsec$ &  0.12 &    20 &      0.60 $\pm$     0.37 &     10.42 $\pm$     0.83 &   0.20 $\pm$  0.22 &  $-$0.13 $\pm$  0.04 &   0.12 &   0.96 &  2.31E-01 \\
   12 & w & C3 &   9$^h$12$^m$00.85$^s$ & $-$64$\degr$51$\arcmin$58.04$\arcsec$ &  0.14 &    15 &      0.43 $\pm$     0.30 &      1.56 $\pm$     0.45 &   0.07 $\pm$  0.26 &   0.11 $\pm$  0.16 &   0.12 &   1.22 &  7.19E-01 \\
   13 & w & C4 &   9$^h$12$^m$03.53$^s$ & $-$64$\degr$52$\arcmin$01.77$\arcsec$ &  0.09 &    33 &      0.81 $\pm$     0.39 &      1.37 $\pm$     0.46 &  $-$0.09 $\pm$  0.17 &  $-$0.18 $\pm$  0.15 &   0.12 &   1.48 &  5.80E-01 \\
   14 & h & C4 &   9$^h$12$^m$03.89$^s$ & $-$64$\degr$52$\arcmin$03.16$\arcsec$ &  0.28 &     4 &      0.09 $\pm$     0.07 &      1.37 $\pm$     0.46 &  $-$0.50 $\pm$  0.43 &  $-$0.18 $\pm$  0.15 &   0.12 &   1.88 &  9.26E-01 \\
   15 & w & -- &   9$^h$12$^m$02.46$^s$ & $-$64$\degr$52$\arcmin$06.23$\arcsec$ &  0.12 &    20 &      0.44 $\pm$     0.26 &       --  &  $-$0.40 $\pm$  0.20 &     --   &   0.12 &   1.87 &  7.56E-01 \\
   16 & w & -- &   9$^h$12$^m$04.62$^s$ & $-$64$\degr$52$\arcmin$09.79$\arcsec$ &  0.05 &    97 &      2.40 $\pm$     0.66 &       --  &  $-$0.15 $\pm$  0.10 &     --   &   0.12 &   1.55 &  1.85E-03 \\
   17 & w & -- &   9$^h$12$^m$00.96$^s$ & $-$64$\degr$50$\arcmin$54.48$\arcsec$ &  0.21 &     7 &      0.12 $\pm$     0.09 &       --  &  $-$0.43 $\pm$  0.34 &     --   &   0.12 &   2.57 &  8.32E-03 \\
   18 & w & -- &   9$^h$11$^m$53.61$^s$ & $-$64$\degr$52$\arcmin$06.42$\arcsec$ &  0.23 &     6 &      0.23 $\pm$     0.17 &       --  &  -1.00 $\pm$  0.26 &     --   &   0.12 &   2.67 &  3.57E-01 \\
   19 & w & -- &   9$^h$12$^m$14.80$^s$ & $-$64$\degr$51$\arcmin$41.96$\arcsec$ &  0.21 &     8 &      0.25 $\pm$     0.24 &       --  &   0.00 $\pm$  0.35 &     --   &   0.12 &   0.89 &  4.18E-01 \\
   20 & w & -- &   9$^h$11$^m$48.86$^s$ & $-$64$\degr$52$\arcmin$49.45$\arcsec$ &  0.27 &     5 &      0.10 $\pm$     0.07 &       --  &  -1.00 $\pm$  0.30 &     --   &   0.12 &   3.20 &  9.68E-01 \\
   21 & w & -- &   9$^h$11$^m$44.48$^s$ & $-$64$\degr$51$\arcmin$28.20$\arcsec$ &  0.12 &    27 &      0.65 $\pm$     0.55 &       --  &  $-$0.26 $\pm$  0.19 &     --   &   0.12 &   1.55 &  5.05E-02 \\
   22 & w & -- &   9$^h$12$^m$17.46$^s$ & $-$64$\degr$53$\arcmin$09.41$\arcsec$ &  0.29 &     5 &      0.13 $\pm$     0.12 &       --  &  $-$0.20 $\pm$  0.44 &     --   &   0.12 &   1.32 &  9.37E-01 \\
   23 & w & -- &   9$^h$12$^m$08.06$^s$ & $-$64$\degr$49$\arcmin$40.57$\arcsec$ &  0.16 &    20 &      0.60 $\pm$     0.45 &       --  &  $-$0.20 $\pm$  0.22 &     --   &   0.12 &   1.13 &  8.82E-02 \\
   24 & w & -- &   9$^h$11$^m$51.14$^s$ & $-$64$\degr$49$\arcmin$43.26$\arcsec$ &  0.21 &    11 &      0.48 $\pm$     0.29 &       --  &   0.45 $\pm$  0.27 &     --   &   0.12 &   0.29 &  9.65E-01 \\
   25 & w & -- &   9$^h$11$^m$41.79$^s$ & $-$64$\degr$50$\arcmin$28.62$\arcsec$ &  0.26 &     8 &      0.85 $\pm$     0.79 &       --  &   0.50 $\pm$  0.31 &     --   &   0.12 &  $-$0.31 &  5.39E-01 \\
   26 & w &  46 &   9$^h$12$^m$26.51$^s$ & $-$64$\degr$52$\arcmin$57.02$\arcsec$ &  0.17 &    21 &      0.45 $\pm$     0.29 &      0.66 $\pm$     0.27 &  $-$0.62 $\pm$  0.17 &  $-$0.41 $\pm$  0.16 &   0.12 &   2.41 &  1.63E-01 \\
   27 & w & -- &   9$^h$12$^m$04.15$^s$ & $-$64$\degr$48$\arcmin$55.47$\arcsec$ &  0.20 &    17 &      0.86 $\pm$     0.54 &       --  &   0.53 $\pm$  0.21 &     --   &   0.12 &  $-$0.18 &  3.06E-01 \\
   28 & h & -- &   9$^h$11$^m$35.62$^s$ & $-$64$\degr$51$\arcmin$59.52$\arcsec$ &  0.26 &     8 &      0.44 $\pm$     0.33 &       --  &   0.50 $\pm$  0.31 &     --   &   0.12 &  $-$0.28 &  8.31E-01 \\
   29 & w & -- &   9$^h$11$^m$39.41$^s$ & $-$64$\degr$50$\arcmin$01.65$\arcsec$ &  0.30 &     8 &      1.37 $\pm$     0.69 &       --  &   0.50 $\pm$  0.31 &     --   &   0.12 &  $-$0.70 &  2.84E-01 \\
   30 & w & -- &   9$^h$11$^m$33.32$^s$ & $-$64$\degr$51$\arcmin$34.27$\arcsec$ &  0.33 &     6 &      0.16 $\pm$     0.16 &       --  &   0.33 $\pm$  0.38 &     --   &   0.12 &   0.97 &  9.36E-01 \\
   31 & w &  17 &   9$^h$11$^m$33.32$^s$ & $-$64$\degr$51$\arcmin$03.25$\arcsec$ &  0.12 &    43 &      0.96 $\pm$     0.25 &      1.66 $\pm$     0.42 &  $-$0.91 $\pm$  0.06 &  $-$0.37 $\pm$  0.11 &   0.12 &   3.56 &  8.67E-02 \\
   32 & w & -- &   9$^h$11$^m$50.44$^s$ & $-$64$\degr$48$\arcmin$43.59$\arcsec$ &  0.20 &    21 &      0.45 $\pm$     0.25 &       --  &  $-$0.52 $\pm$  0.19 &     --   &   0.12 &   2.09 &  8.34E-01 \\
   33 & w &  29 &   9$^h$12$^m$02.10$^s$ & $-$64$\degr$55$\arcmin$06.43$\arcsec$ &  0.12 &    39 &      1.00 $\pm$     0.43 &      1.22 $\pm$     0.33 &  $-$0.13 $\pm$  0.16 &  $-$0.22 $\pm$  0.12 &   0.12 &   1.48 &  6.89E-01 \\
   34 & w & -- &   9$^h$12$^m$24.38$^s$ & $-$64$\degr$49$\arcmin$22.59$\arcsec$ &  0.41 &     7 &      0.25 $\pm$     0.24 &       --  &  $-$0.14 $\pm$  0.37 &     --   &   0.12 &   0.84 &  9.84E-01 \\
   35 & w & -- &   9$^h$11$^m$49.00$^s$ & $-$64$\degr$54$\arcmin$48.78$\arcsec$ &  0.27 &     8 &      0.20 $\pm$     0.18 &       --  &   0.00 $\pm$  0.35 &     --   &   0.12 &   1.42 &  5.22E-01 \\
   36 & w &   7 &   9$^h$12$^m$18.58$^s$ & $-$64$\degr$48$\arcmin$41.13$\arcsec$ &  0.10 &   117 &      2.75 $\pm$     0.69 &      2.90 $\pm$     0.49 &  $-$0.35 $\pm$  0.09 &  $-$0.42 $\pm$  0.07 &   0.12 &   1.76 &  7.57E-01 \\
   37 & w &  37 &   9$^h$12$^m$33.48$^s$ & $-$64$\degr$50$\arcmin$27.83$\arcsec$ &  0.13 &    60 &      1.90 $\pm$     0.78 &      1.46 $\pm$     0.40 &   0.03 $\pm$  0.13 &   0.18 $\pm$  0.14 &   0.12 &   1.05 &  4.56E-01 \\
   38 & h & -- &   9$^h$12$^m$21.80$^s$ & $-$64$\degr$54$\arcmin$53.17$\arcsec$ &  0.37 &     7 &      0.26 $\pm$     0.16 &       --  &  -1.00 $\pm$  0.23 &     --   &   0.12 &   3.62 &  8.16E-01 \\
   39 & w &  22 &   9$^h$12$^m$35.07$^s$ & $-$64$\degr$53$\arcmin$18.59$\arcsec$ &  0.16 &    39 &      1.23 $\pm$     0.34 &      0.89 $\pm$     0.25 &  $-$0.90 $\pm$  0.07 &  $-$0.77 $\pm$  0.08 &   0.12 &   4.67 &  6.82E-01 \\
   40 & w & -- &   9$^h$12$^m$15.21$^s$ & $-$64$\degr$55$\arcmin$20.70$\arcsec$ &  0.32 &     9 &      0.47 $\pm$     0.46 &       --  &   0.78 $\pm$  0.21 &     --   &   0.12 &   0.07 &  9.49E-01 \\
   41 & w &   2 &   9$^h$11$^m$28.05$^s$ & $-$64$\degr$50$\arcmin$35.65$\arcsec$ &  0.04 &   513 &     11.69 $\pm$     1.18 &     14.36 $\pm$     1.02 &  $-$0.43 $\pm$  0.04 &  $-$0.30 $\pm$  0.03 &   0.12 &   1.95 &  7.37E-01 \\
   42 & w &  26 &   9$^h$12$^m$27.07$^s$ & $-$64$\degr$48$\arcmin$49.05$\arcsec$ &  0.17 &    55 &      1.53 $\pm$     0.40 &      1.25 $\pm$     0.39 &  $-$0.27 $\pm$  0.13 &  $-$0.35 $\pm$  0.13 &   0.12 &   1.36 &  3.27E-01 \\
   43 & w &  41 &   9$^h$12$^m$34.69$^s$ & $-$64$\degr$53$\arcmin$49.51$\arcsec$ &  0.20 &    29 &      0.72 $\pm$     0.26 &      0.43 $\pm$     0.19 &  $-$0.93 $\pm$  0.07 &  $-$0.85 $\pm$  0.12 &   0.12 &   3.71 &  9.59E-01 \\
   44 & h &  91 &   9$^h$12$^m$30.89$^s$ & $-$64$\degr$54$\arcmin$26.59$\arcsec$ &  0.46 &     6 &      0.36 $\pm$     0.27 &      0.91 $\pm$     0.39 &   1.00 $\pm$  0.26 &   0.57 $\pm$  0.23 &   0.12 &  $-$0.55 &  2.77E-01 \\
   45 & w & -- &   9$^h$12$^m$21.32$^s$ & $-$64$\degr$48$\arcmin$17.39$\arcsec$ &  0.35 &    14 &      0.43 $\pm$     0.31 &       --  &   0.14 $\pm$  0.26 &     --   &   0.12 &   0.92 &  2.10E-01 \\
   46 & h & -- &   9$^h$11$^m$24.70$^s$ & $-$64$\degr$51$\arcmin$56.19$\arcsec$ &  0.36 &     8 &      0.29 $\pm$     0.28 &       --  &   0.25 $\pm$  0.34 &     --   &   0.12 &   0.68 &  6.00E-01 \\
   47 & w &  13 &   9$^h$11$^m$26.46$^s$ & $-$64$\degr$53$\arcmin$14.20$\arcsec$ &  0.13 &    57 &      2.56 $\pm$     1.20 &      4.32 $\pm$     0.66 &   0.40 $\pm$  0.12 &   0.54 $\pm$  0.08 &   0.12 &   0.31 &  6.66E-02 \\
   48 & w &  85 &   9$^h$12$^m$13.85$^s$ & $-$64$\degr$47$\arcmin$51.40$\arcsec$ &  0.29 &    19 &      0.56 $\pm$     0.40 &      0.68 $\pm$     0.32 &   0.05 $\pm$  0.23 &   0.53 $\pm$  0.43 &   0.12 &   1.06 &  8.28E-02 \\
   49 & w & -- &   9$^h$12$^m$30.56$^s$ & $-$64$\degr$48$\arcmin$54.15$\arcsec$ &  0.43 &    11 &      0.77 $\pm$     0.75 &       --  &   0.82 $\pm$  0.17 &     --   &   0.12 &  $-$0.89 &  4.52E-01 \\
   50 & w & -- &   9$^h$12$^m$34.25$^s$ & $-$64$\degr$49$\arcmin$08.26$\arcsec$ &  0.49 &     9 &      0.38 $\pm$     0.30 &       --  &   0.33 $\pm$  0.31 &     --   &   0.12 &   0.10 &  5.12E-01 \\
   51 & w &  50 &   9$^h$12$^m$39.52$^s$ & $-$64$\degr$53$\arcmin$31.60$\arcsec$ &  0.36 &    11 &      0.37 $\pm$     0.20 &      0.51 $\pm$     0.23 &  -1.00 $\pm$  0.16 &  $-$0.67 $\pm$  0.14 &   0.12 &   4.80 &  7.39E-03 \\
   52 & w &  67 &   9$^h$12$^m$42.46$^s$ & $-$64$\degr$50$\arcmin$34.62$\arcsec$ &  0.34 &    16 &      0.38 $\pm$     0.30 &      0.70 $\pm$     0.31 &   0.00 $\pm$  0.25 &  $-$0.17 $\pm$  0.20 &   0.12 &   1.55 &  1.84E-01 \\
   53 & w &  24 &   9$^h$12$^m$01.37$^s$ & $-$64$\degr$56$\arcmin$11.87$\arcsec$ &  0.20 &    27 &      0.74 $\pm$     0.25 &      0.63 $\pm$     0.20 &  $-$0.93 $\pm$  0.07 &  $-$0.91 $\pm$  0.08 &   0.12 &   4.43 &  7.25E-01 \\
   54 & h & -- &   9$^h$11$^m$43.90$^s$ & $-$64$\degr$55$\arcmin$46.53$\arcsec$ &  0.42 &     7 &      0.25 $\pm$     0.24 &       --  &   0.43 $\pm$  0.34 &     --   &   0.12 &   0.67 &  8.73E-01 \\
   55 & w & -- &   9$^h$11$^m$43.64$^s$ & $-$64$\degr$55$\arcmin$55.84$\arcsec$ &  0.34 &    11 &      0.52 $\pm$     0.43 &       --  &   0.64 $\pm$  0.23 &     --   &   0.12 &   0.00 &  1.55E-01 \\
   56 & w &   3 &   9$^h$11$^m$18.86$^s$ & $-$64$\degr$50$\arcmin$58.19$\arcsec$ &  0.06 &   453 &      9.62 $\pm$     1.10 &      9.17 $\pm$     0.77 &  $-$0.59 $\pm$  0.04 &  $-$0.55 $\pm$  0.03 &   0.12 &   2.55 &  3.36E-14 \\
   57 & w & -- &   9$^h$11$^m$17.27$^s$ & $-$64$\degr$52$\arcmin$45.02$\arcsec$ &  0.36 &    13 &      0.27 $\pm$     0.23 &       --  &  $-$0.08 $\pm$  0.28 &     --   &   0.12 &   1.89 &  4.99E-01 \\
   58 & w & -- &   9$^h$11$^m$22.95$^s$ & $-$64$\degr$54$\arcmin$18.78$\arcsec$ &  0.26 &    26 &      0.53 $\pm$     0.27 &       --  &  $-$0.62 $\pm$  0.15 &     --   &   0.12 &   2.66 &  2.18E-02 \\
   59 & h & -- &   9$^h$11$^m$59.68$^s$ & $-$64$\degr$56$\arcmin$46.54$\arcsec$ &  0.43 &     9 &      0.32 $\pm$     0.31 &       --  &  $-$0.11 $\pm$  0.33 &     --   &   0.12 &   0.80 &  4.00E-01 \\
   60 & w & -- &   9$^h$11$^m$15.90$^s$ & $-$64$\degr$52$\arcmin$27.63$\arcsec$ &  0.49 &     8 &      0.17 $\pm$     0.12 &       --  &  $-$0.75 $\pm$  0.23 &     --   &   0.12 &   3.03 &  1.65E-01 \\
   61 & w &  25 &   9$^h$11$^m$28.81$^s$ & $-$64$\degr$55$\arcmin$21.77$\arcsec$ &  0.15 &    79 &      1.74 $\pm$     0.46 &      0.86 $\pm$     0.27 &  $-$0.49 $\pm$  0.10 &  $-$0.79 $\pm$  0.09 &   0.12 &   2.33 &  8.45E-02 \\
   62 & w &  45 &   9$^h$11$^m$34.51$^s$ & $-$64$\degr$55$\arcmin$57.17$\arcsec$ &  0.25 &    28 &      0.89 $\pm$     0.42 &      0.87 $\pm$     0.32 &   0.07 $\pm$  0.19 &  $-$0.43 $\pm$  0.14 &   0.12 &   1.03 &  3.43E-01 \\
   63 & w & -- &   9$^h$12$^m$50.76$^s$ & $-$64$\degr$52$\arcmin$25.27$\arcsec$ &  0.38 &    19 &      0.45 $\pm$     0.18 &       --  &  $-$0.68 $\pm$  0.17 &     --   &   0.12 &   3.21 &  5.90E-02 \\
   64 & w &  62 &   9$^h$12$^m$51.77$^s$ & $-$64$\degr$52$\arcmin$08.18$\arcsec$ &  0.36 &    21 &      0.56 $\pm$     0.36 &      0.70 $\pm$     0.32 &  $-$0.14 $\pm$  0.22 &   0.24 $\pm$  0.31 &   0.12 &   1.55 &  4.21E-01 \\
   65 & w & -- &   9$^h$12$^m$47.28$^s$ & $-$64$\degr$54$\arcmin$06.97$\arcsec$ &  0.30 &    32 &      0.86 $\pm$     0.46 &       --  &  $-$0.25 $\pm$  0.17 &     --   &   0.12 &   1.32 &  4.49E-01 \\
   66 & h & -- &   9$^h$12$^m$06.37$^s$ & $-$64$\degr$57$\arcmin$04.47$\arcsec$ &  0.60 &     6 &      0.36 $\pm$     0.23 &       --  &   0.67 $\pm$  0.30 &     --   &   0.12 &  $-$0.45 &  6.71E-01 \\
   67 & h & -- &   9$^h$12$^m$18.64$^s$ & $-$64$\degr$56$\arcmin$52.58$\arcsec$ &  0.50 &    10 &      0.27 $\pm$     0.22 &       --  &   0.00 $\pm$  0.32 &     --   &   0.12 &   1.00 &  1.24E-01 \\
   68 & w & -- &   9$^h$11$^m$16.26$^s$ & $-$64$\degr$49$\arcmin$32.16$\arcsec$ &  0.61 &     9 &      0.12 $\pm$     0.11 &       --  &   0.33 $\pm$  0.31 &     --   &   0.12 &   1.01 &  4.69E-01 \\
   69 & w & -- &   9$^h$11$^m$53.33$^s$ & $-$64$\degr$57$\arcmin$16.17$\arcsec$ &  0.44 &    12 &      0.55 $\pm$     0.27 &       --  &   0.50 $\pm$  0.25 &     --   &   0.12 &   0.12 &  7.67E-01 \\
   70 & w & -- &   9$^h$11$^m$58.93$^s$ & $-$64$\degr$57$\arcmin$22.85$\arcsec$ &  0.54 &     9 &      0.18 $\pm$     0.13 &       --  &  $-$0.78 $\pm$  0.21 &     --   &   0.12 &   2.83 &  7.86E-01 \\
   71 & w & -- &   9$^h$12$^m$50.21$^s$ & $-$64$\degr$54$\arcmin$19.80$\arcsec$ &  0.51 &    15 &      0.29 $\pm$     0.22 &       --  &   0.20 $\pm$  0.25 &     --   &   0.12 &   1.10 &  9.79E-01 \\
   72 & w &  59 &   9$^h$11$^m$08.50$^s$ & $-$64$\degr$51$\arcmin$13.21$\arcsec$ &  0.40 &    21 &      0.80 $\pm$     0.45 &      0.82 $\pm$     0.39 &   0.14 $\pm$  0.22 &   0.10 $\pm$  0.57 &   0.12 &   0.50 &  1.36E-01 \\
   73 & w &   5 &   9$^h$12$^m$56.91$^s$ & $-$64$\degr$52$\arcmin$27.62$\arcsec$ &  0.10 &   336 &     19.35 $\pm$     3.55 &     19.24 $\pm$     2.08 &   0.65 $\pm$  0.04 &   0.85 $\pm$  0.04 &   0.12 &  $-$0.13 &  5.82E-01 \\
   74 & w &  77 &   9$^h$12$^m$23.34$^s$ & $-$64$\degr$57$\arcmin$10.49$\arcsec$ &  0.36 &    26 &      0.66 $\pm$     0.25 &      0.34 $\pm$     0.27 &  $-$0.92 $\pm$  0.08 &  $-$0.62 $\pm$  0.27 &   0.12 &   3.97 &  5.01E-01 \\
   75 & h & -- &   9$^h$12$^m$31.11$^s$ & $-$64$\degr$57$\arcmin$13.75$\arcsec$ &  0.60 &    15 &      0.64 $\pm$     0.63 &       --  &   0.33 $\pm$  0.24 &     --   &   0.12 &  $-$0.26 &  3.77E-01 \\
   76 & w &  47 &   9$^h$11$^m$12.12$^s$ & $-$64$\degr$55$\arcmin$01.03$\arcsec$ &  0.26 &    58 &      1.49 $\pm$     0.53 &      1.47 $\pm$     0.49 &  $-$0.31 $\pm$  0.12 &  $-$0.05 $\pm$  0.15 &   0.12 &   1.65 &  8.50E-01 \\
   77 & w &   6 &   9$^h$13$^m$01.44$^s$ & $-$64$\degr$52$\arcmin$16.44$\arcsec$ &  0.18 &   146 &      3.74 $\pm$     1.07 &      3.91 $\pm$     0.60 &  $-$0.32 $\pm$  0.08 &  $-$0.30 $\pm$  0.06 &   0.12 &   1.76 &  9.26E-01 \\
   78 & w & -- &   9$^h$11$^m$29.21$^s$ & $-$64$\degr$57$\arcmin$09.05$\arcsec$ &  0.58 &    14 &      0.26 $\pm$     0.23 &       --  &  $-$0.14 $\pm$  0.26 &     --   &   0.12 &   1.75 &  9.06E-01 \\
   79 & w &  86 &   9$^h$12$^m$29.44$^s$ & $-$64$\degr$45$\arcmin$58.22$\arcsec$ &  0.47 &    32 &      2.69 $\pm$     1.22 &      1.18 $\pm$     0.48 &   0.94 $\pm$  0.06 &   1.00 $\pm$  0.15 &   0.12 &  -1.32 &  5.87E-02 \\
   80 & w & -- &   9$^h$12$^m$20.92$^s$ & $-$64$\degr$58$\arcmin$08.86$\arcsec$ &  0.51 &    21 &      0.74 $\pm$     0.67 &       --  &   0.14 $\pm$  0.22 &     --   &   0.12 &   0.57 &  2.06E-01 \\
   81 & w &  88 &   9$^h$11$^m$21.84$^s$ & $-$64$\degr$56$\arcmin$54.49$\arcsec$ &  0.45 &    25 &      1.22 $\pm$     0.72 &      1.20 $\pm$     0.45 &   0.60 $\pm$  0.16 &   0.84 $\pm$  0.17 &   0.12 &  $-$0.30 &  9.69E-03 \\
   82 & w &  14 &   9$^h$12$^m$50.13$^s$ & $-$64$\degr$56$\arcmin$12.30$\arcsec$ &  0.19 &   158 &      4.65 $\pm$     0.97 &      3.05 $\pm$     0.59 &  $-$0.20 $\pm$  0.08 &   0.09 $\pm$  0.09 &   0.12 &   1.45 &  3.55E-02 \\
   83 & w &  38 &   9$^h$11$^m$17.22$^s$ & $-$64$\degr$46$\arcmin$52.99$\arcsec$ &  0.37 &    57 &      1.59 $\pm$     0.59 &      1.54 $\pm$     0.58 &  $-$0.09 $\pm$  0.13 &  $-$0.21 $\pm$  0.15 &   0.12 &   1.16 &  6.94E-01 \\
   84 & w &  23 &   9$^h$12$^m$36.86$^s$ & $-$64$\degr$45$\arcmin$45.01$\arcsec$ &  0.34 &    79 &      1.99 $\pm$     0.64 &      2.39 $\pm$     0.59 &  $-$0.19 $\pm$  0.11 &  $-$0.01 $\pm$  0.11 &   0.12 &   1.62 &  3.68E-01 \\
   85 & w &  19 &   9$^h$10$^m$57.93$^s$ & $-$64$\degr$49$\arcmin$33.42$\arcsec$ &  0.24 &   123 &      2.77 $\pm$     0.88 &      3.26 $\pm$     0.68 &  $-$0.51 $\pm$  0.08 &  $-$0.33 $\pm$  0.08 &   0.12 &   2.10 &  7.64E-02 \\
   86 & w &  76 &   9$^h$12$^m$13.55$^s$ & $-$64$\degr$44$\arcmin$32.76$\arcsec$ &  0.51 &    41 &      0.91 $\pm$     0.51 &      1.32 $\pm$     0.51 &  $-$0.02 $\pm$  0.16 &   0.72 $\pm$  0.19 &   0.12 &   1.32 &  8.39E-01 \\
   87 & h &  56 &   9$^h$12$^m$43.92$^s$ & $-$64$\degr$57$\arcmin$42.63$\arcsec$ &  0.65 &    25 &      0.35 $\pm$     0.20 &      0.64 $\pm$     0.31 &  $-$0.04 $\pm$  0.20 &  $-$0.49 $\pm$  0.19 &   0.12 &   3.04 &  5.64E-01 \\
   88 & w & -- &   9$^h$12$^m$52.44$^s$ & $-$64$\degr$56$\arcmin$55.44$\arcsec$ &  0.53 &    36 &      0.91 $\pm$     0.57 &       --  &  $-$0.17 $\pm$  0.16 &     --   &   0.12 &   1.33 &  1.28E-01 \\
   89 & w & -- &   9$^h$13$^m$05.36$^s$ & $-$64$\degr$54$\arcmin$57.27$\arcsec$ &  0.76 &    18 &      0.29 $\pm$     0.19 &       --  &   0.00 $\pm$  0.24 &     --   &   0.12 &   2.80 &  2.68E-01 \\
   90 & h & -- &   9$^h$12$^m$23.50$^s$ & $-$64$\degr$44$\arcmin$43.09$\arcsec$ &  0.64 &    28 &      0.97 $\pm$     0.49 &       --  &   0.36 $\pm$  0.18 &     --   &   0.12 &   0.65 &  4.08E-01 \\
   91 & w &  53 &   9$^h$11$^m$27.67$^s$ & $-$64$\degr$45$\arcmin$15.12$\arcsec$ &  0.46 &    51 &      1.03 $\pm$     0.46 &      1.55 $\pm$     0.58 &  $-$0.29 $\pm$  0.13 &   0.02 $\pm$  0.21 &   0.12 &   1.94 &  6.07E-01 \\
   92 & h &  31 &   9$^h$13$^m$12.54$^s$ & $-$64$\degr$52$\arcmin$57.87$\arcsec$ &  0.49 &    42 &      1.27 $\pm$     0.75 &      0.96 $\pm$     0.37 &   0.05 $\pm$  0.15 &  $-$0.50 $\pm$  0.15 &   0.12 &   1.18 &  6.04E-01 \\
   93 & w & -- &   9$^h$11$^m$39.48$^s$ & $-$64$\degr$44$\arcmin$36.23$\arcsec$ &  0.77 &    21 &      0.31 $\pm$     0.28 &       --  &   0.05 $\pm$  0.22 &     --   &   0.12 &   0.57 &  9.32E-01 \\
   94 & h &  68 &   9$^h$13$^m$12.10$^s$ & $-$64$\degr$49$\arcmin$13.44$\arcsec$ &  0.68 &    32 &      0.78 $\pm$     0.38 &      0.66 $\pm$     0.38 &  $-$0.12 $\pm$  0.18 &  $-$0.41 $\pm$  0.22 &   0.12 &   1.63 &  9.65E-01 \\
   95 & w &  55 &   9$^h$10$^m$49.56$^s$ & $-$64$\degr$53$\arcmin$01.46$\arcsec$ &  0.52 &    34 &      0.75 $\pm$     0.33 &      1.04 $\pm$     0.46 &  $-$0.35 $\pm$  0.16 &  $-$0.24 $\pm$  0.18 &   0.12 &   2.72 &  3.72E-01 \\
   96 & h & -- &   9$^h$10$^m$57.94$^s$ & $-$64$\degr$55$\arcmin$44.48$\arcsec$ &  0.65 &    26 &      0.47 $\pm$     0.27 &       --  &  $-$0.15 $\pm$  0.19 &     --   &   0.12 &   1.16 &  2.17E-02 \\
   97 & w &  12 &   9$^h$11$^m$37.60$^s$ & $-$64$\degr$44$\arcmin$04.10$\arcsec$ &  0.35 &   112 &      2.66 $\pm$     0.62 &      3.76 $\pm$     0.74 &  $-$0.29 $\pm$  0.09 &  $-$0.27 $\pm$  0.08 &   0.12 &   1.82 &  3.42E-02 \\
   98 & h &  43 &   9$^h$12$^m$09.52$^s$ & -65$\degr$00$\arcmin$14.72$\arcsec$ &  0.62 &    37 &      1.19 $\pm$     0.79 &      1.70 $\pm$     0.53 &   0.14 $\pm$  0.16 &   0.14 $\pm$  0.15 &   0.12 &   0.91 &  6.89E-01 \\
   99 & w &   1 &   9$^h$10$^m$49.68$^s$ & $-$64$\degr$48$\arcmin$14.13$\arcsec$ &  0.25 &   242 &     11.84 $\pm$     3.01 &     55.16 $\pm$     2.59 &   0.35 $\pm$  0.06 &  $-$0.00 $\pm$  0.02 &   0.12 &   0.20 &  2.94E-01 \\
  100 & h &  71 &   9$^h$10$^m$42.82$^s$ & $-$64$\degr$53$\arcmin$13.70$\arcsec$ &  0.71 &    28 &      1.14 $\pm$     0.88 &      1.15 $\pm$     0.44 &   0.21 $\pm$  0.18 &  $-$0.23 $\pm$  0.87 &   0.12 &   0.23 &  1.50E-01 \\
  101 & h &  27 &   9$^h$13$^m$23.50$^s$ & $-$64$\degr$51$\arcmin$04.60$\arcsec$ &  0.71 &    40 &      0.66 $\pm$     0.30 &      2.18 $\pm$     0.59 &   0.05 $\pm$  0.16 &  $-$0.11 $\pm$  0.12 &   0.12 &   1.94 &  9.21E-01 \\
  102 & w & -- &   9$^h$10$^m$48.59$^s$ & $-$64$\degr$55$\arcmin$38.02$\arcsec$ &  0.84 &    24 &      1.01 $\pm$     0.62 &       --  &   0.42 $\pm$  0.19 &     --   &   0.12 &   0.27 &  9.76E-01 \\
  103 & h &  83 &   9$^h$10$^m$39.27$^s$ & $-$64$\degr$51$\arcmin$40.10$\arcsec$ &  0.78 &    30 &      0.50 $\pm$     0.32 &      0.86 $\pm$     0.45 &   0.07 $\pm$  0.18 &  $-$0.06 $\pm$  0.23 &   0.12 &   1.62 &  3.57E-01 \\
  104 & h & -- &   9$^h$12$^m$24.02$^s$ & -65$\degr$00$\arcmin$28.78$\arcsec$ &  0.93 &    26 &      0.51 $\pm$     0.30 &       --  &  $-$0.08 $\pm$  0.20 &     --   &   0.12 &   2.44 &  6.95E-01 \\
  105 & w &  44 &   9$^h$12$^m$10.27$^s$ & $-$64$\degr$42$\arcmin$44.74$\arcsec$ &  0.35 &   168 &      4.21 $\pm$     1.00 &      1.06 $\pm$     0.42 &  $-$0.32 $\pm$  0.07 &  $-$0.60 $\pm$  0.13 &   0.12 &   1.77 &  5.63E-01 \\
  106 & h & -- &   9$^h$10$^m$42.92$^s$ & $-$64$\degr$55$\arcmin$07.85$\arcsec$ &  0.67 &    40 &      0.66 $\pm$     0.34 &       --  &  $-$0.15 $\pm$  0.16 &     --   &   0.12 &   2.62 &  8.54E-02 \\
  107 & w &  16 &   9$^h$13$^m$37.06$^s$ & $-$64$\degr$52$\arcmin$01.10$\arcsec$ &  0.43 &   167 &      6.95 $\pm$     1.67 &      4.99 $\pm$     0.89 &   0.17 $\pm$  0.08 &   0.33 $\pm$  0.09 &   0.12 &   0.58 &  9.68E-01 \\
  108 & h &  61 &   9$^h$11$^m$54.18$^s$ & -65$\degr$02$\arcmin$02.85$\arcsec$ &  1.07 &    30 &      0.42 $\pm$     0.24 &      1.62 $\pm$     0.56 &   0.20 $\pm$  0.18 &   0.48 $\pm$  0.21 &   0.12 &   1.73 &  5.80E-01 \\
  109 & h &  28 &   9$^h$11$^m$57.39$^s$ & -65$\degr$02$\arcmin$52.76$\arcsec$ &  0.85 &    60 &      1.11 $\pm$     0.49 &      1.38 $\pm$     0.45 &  $-$0.03 $\pm$  0.13 &  $-$0.60 $\pm$  0.12 &   0.12 &   1.64 &  2.85E-01 \\
  110 & h &  15 &   9$^h$12$^m$51.60$^s$ & -65$\degr$02$\arcmin$13.39$\arcsec$ &  0.54 &   207 &      4.92 $\pm$     0.94 &      3.80 $\pm$     0.77 &  $-$0.29 $\pm$  0.07 &  $-$0.23 $\pm$  0.09 &   0.12 &   1.53 &  8.32E-01 \\
  111 & h &  20 &   9$^h$13$^m$28.39$^s$ & $-$64$\degr$59$\arcmin$18.40$\arcsec$ &  0.61 &   170 &      1.75 $\pm$     0.36 &      1.47 $\pm$     0.42 &  $-$0.44 $\pm$  0.07 &  $-$0.88 $\pm$  0.07 &   0.12 &   2.79 &  4.60E-01 \\
  112 & h &  90 &   9$^h$14$^m$02.25$^s$ & $-$64$\degr$57$\arcmin$08.79$\arcsec$ &  0.80 &   171 &      1.46 $\pm$     0.56 &      1.50 $\pm$     0.66 &  $-$0.01 $\pm$  0.08 &  $-$0.50 $\pm$  0.28 &   0.12 &   1.51 &  6.26E-01 \\
  113 & h &  79 &   9$^h$13$^m$32.22$^s$ & -65$\degr$01$\arcmin$52.75$\arcsec$ &  0.52 &   218 &      2.43 $\pm$     0.48 &      1.90 $\pm$     0.78 &  $-$0.49 $\pm$  0.06 &  $-$0.14 $\pm$  0.25 &   0.12 &   2.29 &  7.11E-08
\end{longtable}
\end{landscape}
}

\end{document}